\documentstyle[12pt,onecol]{article}
\begin{document}
\setlength{\baselineskip}{3ex}
\begin{flushright}
accepted \  Phys. Rev. C  \hspace{3cm} RU-92-07 \\
\end{flushright}
\begin{center}
{\large\bf Self-Similarity and Scaling Behavior in Nuclear Collision.} \\
\vspace{1cm}
S.J.~Lee$^*$ and A.Z.~Mekjian \\
{\em Dept. of Physics and Astronomy, Rutgers University, Piscataway,
New Jersey 08855, U.S.A.}\\
submitted on June 24, 1992 \\
revised on Dec. 16, 1992 \\
\vfill
{\bf Abstract}
\end{center}
\setlength{\baselineskip}{5ex}
\hspace*{30pt}
A simple model is used to explore issues related to self-similarity,
intermittency and scaling behavior in nuclear collisions. Both scaled
factorial moments and power moments are considered in the
investigation of these features. The product mass yields are used to
construct probability distributions and generalized Renyi entropies.
The scaling properties of the resulting Renyi entropies are studied.
Both exponential and power law behavior are found. A discussion of
dimensions associated with these scaling properties is also given. An
analysis of some experimental data is presented.
\vfill
\setlength{\baselineskip}{3ex}
--------------------------- \\
PACS Numbers: \ 24.60.--k, \ 25.70.Np, \ 05.70.--a, \ 05.40.+j  \\
--------------------------- \\
\setlength{\baselineskip}{2ex}
$^*$ Permanent Address: Dept. of Physics, Kyung Hee Univ., Suwon, Korea \\
{\small S.J.L. : \  Phone; (331) 280-2451, \  \
  E-mail; fsjlee@krkhucc1.bitnet \\
 A.Z.M. : \  Phone; (908) 932-2515, \  \
  E-mail; mekjian@ruthep.rutgers.edu }
\pagebreak
\setlength{\baselineskip}{5ex}
\section{Introduction}

\hspace*{30pt}
Medium energy collisions of nuclei offer a useful way of
studying properties of highly excited nuclear matter.
The colliding nuclei produce a system whose density and
temperature are far removed from normal nuclear densities
at zero or low temperatures.
The fragments that evolve from such a collision offer a
useful tool for studying various properties of these systems.
For example, the fragments carry information on the size of
the emitting system \cite{hanbury}
without having to resort to a Hanbury-Brown-Twiss correlation method.
Recent attention has been drawn to the question of intermittency
in the behavior of the distribution of fragments
\cite{ploszaj,fracnuc1,fracnuc2}.

The original idea of studying intermittency in nuclear collisions
came from the work of Bialas and Peschanski \cite{bialas}
who looked at the rapidity distributions of produced particles
in cosmic ray experiments.
They proposed using scaled factorial moments of these rapidity
distribution to study the possible appearance of intermittent
behavior in such collisions and evidence for non-Poissonian
fluctuations.
The ideas of Bialas and Peschanski evolved from work done
in turbulence by Mandelbrot \cite{mandel} and others \cite{turbul}.
The Bialas-Peschanski idea of using factorial moments
was taken up by Ploszajczak and Tucholski \cite{ploszaj}
to discuss intermittency in the fragment distributions of
medium energy heavy ion collisions.
The possibility of seeing intermittent behavior in the
mass yields of heavy ion collisions was also mentioned
in Ref.\cite{mekjian},
for another point of view; in particular a staircase behavior
of the cumulative mass distribution was discussed in terms of flat
regions reflecting missing mass followed by
clustered rises.

The main purpose of this paper is to explore a connection of a
recent model of a fragmentation process with questions related
to self-similarity, non-Poissonian fluctuations, correlations,
intermittency, and scale invariance which may arise in a nuclear
fragmentation. In this paper we will stress the importance of the mass
distribution rather than the cluster distribution studied by others.
The importance of the mass distribution was already noted
in Ref.\cite{mekjian}.
Also we will use the fraction of the mass in clusters of a
given size to obtain probability distributions which in turn will then
be used to obtain generalized Renyi entropies. The generalized Renyi
entropies can be used to obtain associated generalized dimensions
which include the Hausdorff covering dimension, the information
dimension associated with the Shannon information content of the
probability distribution and the correlation dimension related to the
correlation function of a probability distribution. The scaling
properties of the Renyi entropies will be studied along with the
factorial moments used by others.

This paper is divided as follows:
After summarizing our simple fragmentation model briefly,
we will study various features of the cluster and mass distributions
obtained from this model such as their factorial or power
moments in Section 2.
Section 3 contains relations between power moments
and dimensions and generalized Renyi entropies. A discussion of some
recent data is given in Section 4.
Concluding remarks are in Section 5 and details of various ideas
are given in Appendixes.

\section{Probability Distributions in Fragmentation Phenomena}

\hspace*{30pt}
In a previous set of papers \cite{mekjian,xymodel,general,otherx},
we developed an exactly soluble
model for studying fragmentation phenomena.
In this model each partition or fragmentation distribution
is given a weight and properties of the resulting distribution
of fragments are obtained by ensemble averaging the desired
quantity over these distributions with this weighting factor.
The simple weight chosen in Ref.\cite{mekjian,general} is
\begin{eqnarray}
 & & P_A(\vec n, \vec x) = \left[\frac{A!}{Q_A(\vec x)}\right] \prod_{j=1}^A
       \left[\frac{1}{n_j!} \left(\frac{x_j}{j}\right)^{n_j}\right]
   \hspace{2em} {\rm for} \hspace{2em} \sum_{i=1}^A i n_i = A ,  \nonumber \\
 & & P_A(\vec n, \vec x) = 0 \hspace{2em} {\rm for}
          \hspace{2em} \sum_{i=1}^A i n_i \ne A ,
     \label{partwgt}
\end{eqnarray}
where $\vec n = \{n_i\} = (n_1, n_2, ..., n_A)$ is the partition vector
for the fragmentation of $A$ nucleons into $n_1$ free nucleons,
$n_2$ mass 2 clusters, etc. and
$\vec x = (x_1, x_2, ..., x_A)$ with $x_j$ a parameter characterizing
the cluster of size $j$.
The $x_j$'s are related to the underlying physical picture of the
fragmentation process.
For example, in Ref.\cite{general}, the $x_j$ was related to the
fragmentation volume, the quantum volume, a break up temperature,
binding energy, and level density parameters.
The $P_A(\vec n, \vec x)$, in a $A$-dimensional space $\vec n$, satisfies
\begin{eqnarray}
 \sum_{\{n_i\}} P_A(\vec n, \vec x)
    = \sum_{n_1=0}^\infty \sum_{n_2=0}^\infty \cdots \sum_{n_A=0}^\infty
       P_A(\vec n, \vec x) = 1 ,
\end{eqnarray}
where the sum runs over all the possible values of $n_i$.
The condition $P_A(\vec x, \vec n) = 0$ for $\sum_i i n_i \ne A$
restricts the sums over $n_i$'s to the partitions $\vec n$
having $\sum_i i n_i = A$.
The normalization factor $Q_A(\vec x)$ is the partition function
of a canonical ensemble, i.e.,
\begin{eqnarray}
 Q_A(\vec x) = A! \sum_{\{n_i\}_A} \prod_{j=1}^A
       \left[\frac{1}{n_j!} \left(\frac{x_j}{j}\right)^{n_j}\right]
            \label{qaxi}
\end{eqnarray}
where the sum runs over all the possible partitions $\vec n$ with fixed
\begin{eqnarray}
 A = \sum_{j=1}^A j n_j .   \label{aconst}
\end{eqnarray}
We have
\begin{eqnarray}
 Q_0(\vec x) &=& 1 ,    \nonumber \\
 Q_A(\vec x) &=& 0  \hspace{2em} {\rm for} \hspace{2em}  A < 0 .
         \label{qatheta}
\end{eqnarray}
These obvious facts should not be missed when we consider expressions
such as $Q_{A-k}(\vec x)$ (c.f., Eqs.(\ref{nkqax}) -- (\ref{nkcorr})).
For $x_i = x$ for all $i$, (the $x$ model of Ref.\cite{mekjian}),
the partition function $Q_A(\vec x)$ becomes
\begin{eqnarray}
 Q_A(x) = x(x+1)\cdots(x+A-1) \theta(A)
        = \frac{\Gamma(x+A)}{\Gamma(x)} \theta(A) .    \label{qax}
\end{eqnarray}
For the case of $x_1 = xy$ and $x_i = x$ for $i \ge 2$, the $x$-$y$ model
of Ref.\cite{xymodel,general}, the partition function $Q_A(\vec x)$ becomes
\begin{eqnarray}
 Q_A(x,y) = \sum_{r=0}^A \frac{A!}{r!(A-r)!} Q_r(x) \left[x(y-1)\right]^{A-r}
               \theta(A) .   \label{qaxy}
\end{eqnarray}
The step function $\theta(a)$ is zero for $a < 0$ and is one otherwise.
The $\Gamma(x)$ is the gamma function.
The total multiplicity is
\begin{eqnarray}
 M = \sum_{j=1}^A n_j    \label{multp}
\end{eqnarray}
and the $x$ and $y$ are parameters
which depend on thermodynamic variables (the interaction volume
$V$ and the temperature $T$) and on binding energy and level density
parameters.
Details for this case can be found in Refs.\cite{mekjian,general,otherx}.
Here we just mention that a small $x$ corresponds to a low temperature
or a small interaction volume
and a large $x$ corresponds to a high $T$ or a large $V$.

Various properties of the distribution of fragments can be obtained exactly
for the weight of Eq.(\ref{partwgt}).
Most quantities of interest are simply related to the
partition function $Q_A(\vec x)$.
For example, one quantity of interest is the ensemble averaged
mean number of fragments of size $k$ \cite{general}:
\begin{eqnarray}
 <n_k> = \sum_{\{n_i\}_A} n_k P_A(\vec n, \vec x)
       = \frac{x_k}{k} \frac{A!}{(A-k)!} \frac{Q_{A-k}(\vec x)}{Q_A(\vec x)}
    \label{nkqax}
\end{eqnarray}
where $Q_{A-k}(\vec x)$ is also the partition function given by
Eq.(\ref{qaxi}) for $A-k$ nucleons.
Substituting Eq.(\ref{qax}) into Eq.(\ref{nkqax}), for the $x$ model,
the $<n_k>$ is then
\begin{eqnarray}
 <n_k> = \frac{x}{k} \frac{A!}{(A-k)!} \frac{\Gamma(x+A-k)}{\Gamma(x+A)}
                  \label{nkx}
\end{eqnarray}
for $1 \le k \le A$.
The $<n_k>$ satisfies the constraint or sum rule
$\sum_k k <n_k> = A$ which is easy to verify using Eq.(\ref{factxup}).
The expression obtained from this exactly soluble model for
$<n_k>$ gives a very good fit to the inclusive data on
proton-nucleus and nucleus-nucleus collisions \cite{nucfit}.
It exhibits power law behavior, $<n_k> = 1/k$, for $x = 1$, exponential
behavior for large $x > 1$ (high temperature $T$),
and a U shape behavior at small $x < 1$ (low $T$).

This exactly soluble model can also be used to evaluate all
moments of the distribution in closed analytic form \cite{general}
which are as simple as that of Eq.(\ref{nkqax}).
Specifically, using the notation of
\begin{eqnarray}
 \tilde n_i^q \equiv \frac{n_i!}{(n_i-q)!}
     = n_i (n_i - 1) (n_i - 2) \cdots (n_i - q + 1)   \label{factniq}
\end{eqnarray}
for the factorial moment, we obtain
\begin{eqnarray}
 <\tilde n_k^q> = \sum_{\{n_i\}_A} \frac{n_k!}{(n_k-q)!} P_A(\vec n, \vec x)
   = \left[\frac{x_k}{k}\right]^q \frac{A!}{(A-qk)!}
          \frac{Q_{A-qk}(\vec x)}{Q_A(\vec x)} ,
     \label{nkfact}
\end{eqnarray}
and for the various correlations  we have
\begin{eqnarray}
 \left<\tilde n_i^l \tilde n_j^m \tilde n_k^n \cdots \right>
  &=& \left\{ \left[\frac{x_i}{i}\right]^l \left[\frac{x_j}{j}\right]^m
    \left[\frac{x_k}{k}\right]^n \cdots \right\}
      \nonumber \\
  & & \hspace{0.7cm} \times ~ \frac{A!}{(A - \{il + jm + kn + \cdots\})!}
        \frac{Q_{A - \{il + jm + kn + \cdots\}}(\vec x)}{Q_A(\vec x)} .
      \label{nkcorr}
\end{eqnarray}
Eq.(\ref{nkcorr}) is applicable for cases in which any of the $i$, $j$,
$k$, ... are the same index with the following prescription for
the left hand side. If $i = k$, as an illustrative example, then
$<\tilde n_i^l \tilde n_j^m \tilde n_k^n \cdots>$
in the left hand side  is replaced by
$<\tilde n_i^{(l+n)} \tilde n_j^m \cdots>$.

The probability distribution $P_A(\vec n, \vec x)$ of Eq.(\ref{partwgt})
assigns a weight for each partition $\vec n$ normalized
over all partitions subject to the canonical ensemble constraint of
Eq.(\ref{aconst}).
In other words, $P_A(\vec n, \vec x)$ is the probability distribution
on an $(A-1)$-dimensional surface with fixed $A$
in an $A$-dimensional vector space made of partitions
$\vec n = (n_1, n_2, ..., n_A)$.
We can also consider various other probability
distributions by projecting this $P_A(\vec n, \vec x)$ onto
different parts of this vector space.
In subsection 2.1, we consider a probability distribution in a
one-dimensional vector space associated with a single cluster of
a given size $r$ and look at the distribution $P_A(n_r)$,
where $n_r$ is the number of such clusters. We will also compare this
distribution with a Poisson distribution.
In subsection 2.2,
we discuss scaled factorial moments of the cluster distribution.
In subsection 2.3, we study features associated with the fraction of
mass ($kn_k/A$) in cluster size
space and also staircase properties of the cumulative mass distribution
of this fraction.
Subsection 2.4 discusses power moments of the mass distribution.

\subsection{Probability Distribution for Clusters}

\hspace*{30pt}
The probability distribution $P_A(\vec n, \vec x)$ of Eq.(\ref{partwgt})
can be used to obtain
not only the ensemble averaged quantities such as
Eqs.(\ref{nkqax}) -- (\ref{nkcorr}),
but also expressions for the probability of having $n_r$
clusters of size $r$ \cite{general}.
\begin{eqnarray}
 P_A(n_r,\vec x) &=& \sum_{\{n_i\}_{A,n_r}} P_A(\vec n, \vec x)
            \  = \  \left[\frac{A!}{Q_A(\vec x)}\right]
         \left[\frac{1}{n_r!}\left(\frac{x_r}{r}\right)^{n_r}\right]
         \left[\frac{D_{A-n_r r}^{(r)}(\vec x)}{(A - n_r r)!}\right]
              \label{paxinr}
\end{eqnarray}
where the $D$ functions are
\begin{eqnarray}
 D_{A-n_r r}^{(r)}(\vec x) &=& (A-n_r r)! \sum_{\{n_i\}_{A,n_r}} \prod_{j \ne
r}
       \left[\frac{1}{n_j!} \left(\frac{x_j}{j}\right)^{n_j}\right] .
            \label{daxi}
\end{eqnarray}
The sum runs over all the possible partitions $\vec n$
having a fixed $A$ and a fixed number $n_r$ of cluster size $r$.
The product runs in Eq.(\ref{daxi}) over all clusters except the size $r$.
The $P_A(n_r,\vec x)$ also satisfy the following equations.
\begin{eqnarray}
 & & P_A(n_r, \vec x) = 0 \hspace{2em} {\rm for} \hspace{2em} r n_r > A ,
            \nonumber \\
 & & \sum_{n_r = 0}^\infty P_A(n_r, \vec x) = 1 .
\end{eqnarray}
The $D_n^{(r)}(\vec x)$ is also a canonical partition function
for the fragmentation of $n$ nucleons without clusters of size $r$.
Here we do not assume any form for the distributions of these
probabilities $P_A(n_r, \vec x)$ such as a Poisson distribution given by
\begin{eqnarray}
 P_P(n) = \frac{\bar n^n}{n!} e^{-\bar n} ,
   \hspace{2cm} 0 \le n < \infty
  \label{poissd}
\end{eqnarray}
which is normalized to one.
The form of $P_A(n_r,\vec x)$ depends on the parameters $\vec x$
and $A$.
Due to the conservation constraint for $A$,
each cluster in Eqs.(\ref{partwgt}) and
(\ref{paxinr}) is not independent of any other.
For $n_r r$ much smaller than $A$,
\begin{eqnarray}
 \tilde D_{A-n_rr}^{(r)}(\vec x) = D_{A-n_rr}^{(r)}(\vec x)/(A-n_rr)!
             \label{tild}
\end{eqnarray}
is approximately independent of $n_r$. Thus $P_A(n_r,\vec x)$
can be approximated as a Poisson distribution in $n_r$ with
the mean cluster number $<n_r> \approx x_r/r$.

The probability of $n_r$ clusters of size $r$ is given in the
$x$ model by,
(from Eq.(\ref{paxinr}) with $x_i = x$ for all $i$),
\begin{eqnarray}
 P_A(n_r, x) &=& \left[\frac{x}{r}\right]^{n_r} \frac{A!}{n_r! (A - n_r r)!}
            \frac{D_{A-n_r r}^{(r)}(x)}{Q_A(x)}   \nonumber \\
             &=& \left[\frac{A!}{Q_A(x)}\right]
         \left[\frac{1}{n_r!}\left(\frac{x}{r}\right)^{n_r}\right]
         \left[\frac{D_{A-n_r r}^{(r)}(x)}{(A - n_r r)!}\right] .
   \label{panrx}
\end{eqnarray}
In general $D_A^{(r)}$ can be obtained by the recurrence relationship
($y = 0$ case of Eq.(\ref{drrec}) in Appendix B)
\begin{eqnarray}
 D_{n+1}^{(r)}(x) = (n+x) D_n^{(r)}(x) - x \tilde n^{r-1} D_{n-r+1}^{(r)}(x)
     + x \tilde n^r D_{n-r}^{(r)}(x)   \label{drrecc}
\end{eqnarray}
with $\tilde n^r$ given by Eq.(\ref{factniq}).
Obviously, the $D_n^{(r)}(x) = Q_n(x) = x(x+1)\cdots(x+n-1)$ for $n < r$.
The $D_1^{(1)} = 0$,
$D_2^{(1)}(x) = x$, $D_2^{(2)}(x) = x^2$,
$D_3^{(1)}(x) = 2x$, $D_3^{(2)}(x) = x(x^2+2)$,
and $D_3^{(3)}(x) = x^2 (x + 3)$.

As a specific application of the above expression,
we consider the distribution of free nucleons.
Then $r = 1$ and, in the $x$ model,
\begin{eqnarray}
 P_A(n_1,x) = \left[\frac{Q_A(x)}{A!}\right]^{-1}
         \left[\frac{1}{n_1!} x^{n_1}\right]
         \left[\frac{D_{A-n_1}^{(1)}(x)}{(A-n_1)!}\right] .
   \label{pan1x}
\end{eqnarray}
The $D_A(x) \equiv D_A^{(1)}(x)$ is the partition function $Q_A(x, y=0)$
in the $x$-$y$ model of Eq.(\ref{qaxy}) which is studied in detail
in Refs.\cite{general,otherx}.
According to Eqs.(\ref{nkx}) and (\ref{nkfact}),
\begin{eqnarray}
 <n_1> &=& \sum_{n_1 = 0}^\infty n_1 P_A(n_1,x)
             \  = \  \frac{Ax}{A+x-1} ,  \label{n1x} \\
 \sigma^2 &=& <n_1^2> - <n_1>^2 \  = \  <n_1> + <n_1>^2 \frac{(1-x)}{A(A+x-2)}
{}.
          \label{sigmx}
\end{eqnarray}
For $x = 1$, both the mean $<n_1>$ and the fluctuation $\sigma^2$ are 1.
For a Poisson distribution $\sigma^2$ = $<n_1>$. Thus for $x < 1$, the
fluctuation $\sigma$
is larger than Poissonian
($\sigma^2 > \langle n_1\rangle$)
while for $x > 1$, the fluctuation is less than Poissonian
($\sigma^2 < \langle n_1\rangle$).
For $x >> A$,
$\sigma^2 \approx <n_1>${\Large $\left(1 - \frac{<n_1>}{A}\right)$},
while for $x = A$, $<n_1> \approx A/2$ and
$\sigma^2 \approx \frac{3}{4} <n_1>$.
Also at $x=1$, $<n_1> = 1$ and
$P_A(n_1,1) = \frac{1}{n_1!} \sum_{j=0}^{A-n_1} \frac{(-1)^j}{j!}
   \approx \frac{1}{n_1!} e^{-1}$.

The resulting distribution $P_A(n_1,x)$ is illustrated in Fig.1
for several values of the mean number $<n_1>$ of monomers (with
equivalent values of $x$)
and is compared with a Poisson distribution.
Fig.1 shows a plot of $P_A(n_1,x)$ as a function of $n_1$
using Eq.(\ref{pan1x}).
We evaluate $P_A(n_1,x)$ for $A = 200$,
$<n_1> = 7.3823$ which corresponds to $x = 7.6269$ from
Eq.(\ref{n1x}) (solid curve for $k = 1$).
The distribution $P_A(n_1,x)$ is compared with a Poisson
distribution $P_P(n_1)$ given by Eq.(\ref{poissd}).
Two distributions coincide to such an extent that the Poisson
can't be distinguished from the exact result in the figure.
For large $A$, Eq.(\ref{panrx}) reduces to the Poisson distribution
with the mean number of $\bar n_r = x/r$ for any value of $r$.
The extra factors in Eq.(\ref{panrx}) becomes the normalization
factor and is $A$ independent as $A \to \infty$.
The Poissonian behavior in $P_A(n_r,x)$ originates from the same
factor {\Large $\frac{1}{n_r!} \left(\frac{x_r}{r}\right)^{n_r}$}
appearing in the weight in $P_A(\vec n,\vec x)$ of Eq.(\ref{partwgt}).
$P_A(n_r,\vec x)$ deviates from a pure Poisson distribution
because of the constraint of finite $A$.
Without this constraint, $P_A(n_r,\vec x)$ would be a
pure Poisson distribution.
Also shown in Fig.1 (dotted curve) is the case of $A = 26$ with the
same $<n_1> = 7.3823$ and the corresponding $x = 9.9130$
which corresponds to the case considered in Ref.\cite{fracnuc1}. For
this case, departures from a Poisson distribution can be seen.
Note in Fig.1 that the distribution of $k = 2$ and $k = 4$ for
$A = 200$ are also indistinguishable from the Poisson distribution.
At the particular point $x = 1$, and for any size $r$,
we have $<\tilde n_r^q> = <n_r>^q$.
We also note here that for large $<n> = \bar n$, both the $P_A(n_r,x)$
and the Poisson distribution $P_P(n)$ become a
Gaussian distribution.

Similar arguments can be used to obtain the two-dimensional
distribution function for $n_1$ free nucleons and $n_2$ clusters
of size 2 which we call $P_A(n_1,n_2, \vec x)$:
\begin{eqnarray}
 P_A(n_1,n_2,\vec x) &=& \sum_{\{n_i\}_{A,n_1,n_2}} P_A(\vec n, \vec x)
               \nonumber \\
    &=& \left[\frac{Q_A(\vec x)}{A!}\right]^{-1}
        \left[\frac{1}{n_1!} x_1^{n_1}\right]
        \left[\frac{1}{n_2!}\left(\frac{x_2}{2}\right)^{n_2}\right]
      \left[\frac{D_{A-n_1-2n_2}^{(1,2)}(\vec x)}{(A-n_1-2n_2)!}\right]
\end{eqnarray}
with
\begin{eqnarray}
 D_{A-n_1 - 2n_2}^{(1,2)}(\vec x)
 &=& (A-n_1 - 2 n_2)! \sum_{\{n_i\}_{A,n_1,n_2}} \prod_{j \ne 1,2}
       \left[\frac{1}{n_j!} \left(\frac{x_j}{j}\right)^{n_j}\right] .
\end{eqnarray}
For the $x$ model, the recurrence relation is given by (see
Eq.(\ref{d12rrec}) in Appendix B)
\begin{eqnarray}
 D_{n+1}^{(1,2)}(x) = n D_n^{(1,2)}(x) + x n (n-1) D_{n-2}^{(1,2)}(x)
\end{eqnarray}
and $D_0^{(1,2)}(x) = 1$, $D_1^{(1,2)}(x) = 0$, and $D_2^{(1,2)}(x) = 0$.
In general
\begin{eqnarray}
 P_A(n_1,n_2,...,n_r,\vec x)
      &=& \sum_{\{n_i\}_{A,n_1,n_2,...,n_r}} P_A(\vec n,\vec x)
           \nonumber \\
      &=& \left[\frac{Q_A(\vec x)}{A!}\right]^{-1}
      \prod_{k=1}^r\left[\frac{1}{n_k!}\left(\frac{x_k}{k}\right)^{n_k}\right]
          \left[\frac{D_{A-\sum_{j=1}^r j n_j}^{(1,2,...,r)}(\vec x)}
                   {(A - \sum_{j=1}^r j n_j)!}\right]
\end{eqnarray}
with
\begin{eqnarray}
 D_{A - \sum_{j=1}^r j n_j}^{(1,2, ..., r)}(\vec x)
 &=& (A - \sum_{j=1}^r j n_j)! \sum_{\{n_i\}_{A,n_1,n_2,...,n_r}}
      \prod_{j \ne 1,2,...,r}
       \left[\frac{1}{n_j!} \left(\frac{x_j}{j}\right)^{n_j}\right] .
            \label{dax12r}
\end{eqnarray}
For the $x$ model, the recurrence relation that determines the
$D$ functions is given by (see Eq.(\ref{d12rrec}) in Appendix B)
\begin{eqnarray}
 D_{n+1}^{(1,2,...,r)}(x) = n D_n^{(1,2,...,r)}(x)
        + x \tilde n^r D_{n-r}^{(1,2,...,r)}(x)
\end{eqnarray}
and
\begin{eqnarray}
 D_0^{(1,2,...,r)}(x) &=& 1 ,   \nonumber \\
 D_k^{(1,2,...,r)}(x) &=& 0 \hspace{3em} {\rm for} \hspace{3em}
                             1 \le k \le r ,  \nonumber \\
 D_{r+1}^{(1,2,...,r)}(x) &=& x r! .
\end{eqnarray}
When $r \to A$, the $P_A(n_1,n_2,...,n_A, \vec x) = P_A(\vec n, \vec x)$
is just the weight of Eq.(\ref{partwgt}).
The $P_A(n_1,n_2,...,n_r,\vec x)$ satisfies the conditions
\begin{eqnarray}
 && P_A(n_1, n_2, ..., n_r, \vec x) = 0 \hspace{2em}
    {\rm for} \hspace{2em} \sum_{j=1}^r j n_j > A ,  \nonumber \\
 && \sum_{n_1=0}^\infty \sum_{n_2=0}^\infty \cdots \sum_{n_r=0}^\infty
    P_A(n_1, n_2, ..., n_r, \vec x) = 1 .
\end{eqnarray}
Similarly, we can also consider the probability
$P_A(n_s,n_{s+1},...,n_r, \vec x)$.
The recurrence relation for this case is given by Eq.(\ref{dsrrec})
in Appendix B.

\subsection{Scaled Factorial Moments of the Cluster Distribution}

\hspace{30pt}
In this subsection, we consider
the bin size dependences of the scaled factorial moments
which have been proposed as a method for
studying non-statistical fluctuations \cite{bialas} and used
in Refs.\cite{ploszaj,fracnuc1,fracnuc2}.
These moments are obtained as follows.
The physical contents contained in these moments is also discussed.

A length interval $A$ is divided into bins of size $L$,
so that there are $[A/L]$ bins.
Here $[a]$ means the smallest integer not less than $a$, i.e.,
$[a] - 1 < a \le [a]$.
Defining $N_J$ as the number of particles or multiplicity
in bin $J$, with $J$ running from 1 to $[A/L]$,
the $q$'th order scaled factorial moments $F_q$ is defined as
\begin{eqnarray}
 F_q(L) = \frac{\sum_{J=1}^{[A/L]} <N_J (N_J -1) \cdots (N_J - q + 1)>}
                {\sum_{J=1}^{[A/L]} <N_J>^q}
        = \frac{\sum_{J=1}^{[A/L]} <\tilde N_J^q>}
                {\sum_{J=1}^{[A/L]} <N_J>^q} .   \label{fqldef}
\end{eqnarray}
The brackets, $<>$, indicate that the quantity inside it is ensemble
averaged and $\tilde N_J^q$ is the factorial defined by Eq.(\ref{factniq}).
The $<\tilde N_J^q>$ is the ensemble average of the factorial moment
of the total number of clusters $N_J$
in a given bin $J$.
This average is taken over all the possible partitions (events)
with the probability $P_A(\vec n, \vec x)$ given in Eq.(\ref{partwgt}).
For $q = 1$, we can define a cluster distribution probability $P(J,L)$
in a space made of clusters grouped into bins of length $L$ as follow;
\begin{eqnarray}
 P(J,L) &\equiv& \frac{<N_J>}{\sum_{J=1}^{[A/L]} <N_J>}
         = \frac{<N_J>}{<M>}      \nonumber \\
 F_{q=1}(L) &=& \sum_{J=1}^{[A/L]} P(J,L) = 1 .   \label{pjl}
\end{eqnarray}
The $<M>$ is the mean multiplicity given by Eq.(\ref{multp}).
A quantity defined by $P'(J,L) \equiv N_J / <M>$ for each event
is not a probability distribution normalized to one, i.e.,
$\sum_J P'(J,L) \ne 1$. Only the ensemble average of $P'(J,L)$
becomes the probability of Eq.(\ref{pjl}),
$P(J,L) = \left<P'(J,L)\right>$.
In contrast, the mass distribution $P_A(k,\vec x)$ of Eq.(\ref{pakx})
in subsection 2.3 is a probability distribution in an event.

The scaled factorial moment defined by Eq.(\ref{fqldef}) can be
rewritten as
\begin{eqnarray}
 F_q(L) = \sum_{J=1}^{[A/L]}
    \left[\frac{<N_J>^q}{\sum_{J=1}^{[A/L]} <N_J>^q}\right]
    \left[\frac{<\tilde N_J^q>}{<N_J>^q}\right] .  \label{fqlalt}
\end{eqnarray}
Thus the quantity $F_q(L)$ is nothing but a linear combination
(average over bins) of the scaled factorial moments in each bin $J$
(the second square bracket in Eq.(\ref{fqlalt})) with the weight
given by the quantity in the first square bracket.
For a Poisson distribution of $N_J$, $<\tilde N_J^q> = <N_J>^q$
due to Eq.(\ref{factbin}) and thus the $F_q(L) = 1$.
The scaled factorial moment $F_q(L)$ characterizes the distribution
$P_A(\vec n, \vec x)$ compared to the Poisson distribution of
Eq.(\ref{poissd}). Departures from unity have been taken as
evidence for non-Poissonian fluctuations.
Explicitly, for $q = 2$, the scaled factorial moment is
\begin{eqnarray}
 F_2(L) = 1 + \sum_{J=1}^{[A/L]}
    \left[\frac{<N_J>^2}{\sum_{J=1}^{[A/L]} <N_J>^2}\right]
    \left[\frac{\left(<N_J^2> - <N_J>^2\right) - <N_J>}{<N_J>^2}\right]
\end{eqnarray}
and thus becomes 1 for a Poisson distribution $N_J$.
For a bin size $L = 1$, $\left<N_J^2\right> - \left<N_J\right>^2$
is the fluctuation
of the number of clusters of size $J$ in the ensemble.
For the case of bin size $L$ larger than 1,
\begin{eqnarray}
 \frac{\left(<N_J^2> - <N_J>^2\right) - <N_J>}{<N_J>^2}
  &=& \sum_{k \in J} \frac{\left[\left(<n_k^2> - <n_k>^2\right) - <n_k>\right]}
          {<n_k>^2} \frac{<n_k>^2}{<N_J>^2}       \nonumber \\
  & & \hspace{-4.0cm} + \sum_{j \in J} \sum_{k \ne j \in J}
          \frac{\left[<n_j n_k> - <n_j> <n_k>\right]}{<n_j> <n_k>}
          \frac{<n_j> <n_k>}{<N_J>^2} .   \label{njflcor}
\end{eqnarray}
Here $\sum_{k \in J}$ means a summation over $k$ which are in
bin $J$, i.e., $k = (J-1)L + 1$, $(J-1)L + 2$, ..., $JL$.
The quantity in the first square bracket is the departure of
the fluctuation from a Poisson distribution for $n_k$.
The quantity in the second square bracket represents the
departure of the correlation from an independent distribution
of clusters in the same bin $J$
and gives a bin size $L$ dependence.
This correlation departure becomes zero in the $x$ model when
we can neglect the finite $A$ effect such as for a large $x$.
Zero correlation means $<n_j n_k> = <n_j> <n_k>$;
the distribution $n_j$ and the distribution $n_k$ are then independent.
The magnitude of $F_q(L)$ arises from the linearly combined effect
of the two terms in Eq.(\ref{njflcor}), one term related to fluctuations, the
other term related to correlations.
The size $L$ dependence of $F_q(L)$ arises from the correlation term.
Beside an independent Poisson distribution of $n_k$,
$F_q(L)$ can also be unity in a mean field approximation,
i.e., for $<n_j n_k> \approx <n_j> <n_k>$ for
any $j$ and $k$ with large $<n_k>$.

The size $L$ dependence of $F_q(L)$ originates from the
underlying scale or correlation length of order $q$
in Eq.(\ref{njflcor}).
If
\begin{eqnarray}
 F_2(L+1) - F_2(L) \  \propto \  \left<\sum_{|i-j| = L}
       \frac{n_i n_j}{<n_i> <n_j>} \right>  \approx e^{-L/\xi}
\end{eqnarray}
with $\xi$ independent of $L$, then $\xi$ is called the
correlation length of the distribution.
On the other hand, if
\begin{eqnarray}
 F_q(L) \propto L^{\alpha(q)} ,
\end{eqnarray}
then $F_q(L)$ is called scale invariant with the intermittency
exponent $\alpha(q)$.
If the distribution of $n_k$ is either uniform over the whole bin
with $N_J >> q$ or peaked in one cluster size,
then $F_q(L)$ becomes independent of the bin size $L$ with $F_q(L) = 1$.
$F_q(L)$ becomes 1 when the distribution $n_k$ is an independent
Poissonian or when $<N_J>$ is much larger than the order $q$ and
the fluctuation of $N_J$, i.e., in a mean field approximation.

Using the result of Eq.(\ref{nkcorr}), from Eq.(\ref{fqldef}),
\begin{eqnarray}
 F_q(L) &=& \left. \sum_{J=1}^{[A/L]} {\cal N}_J^q(\vec x,L)
            \right/ \sum_{J=1}^{[A/L]} {\cal D}_J^q(\vec x,L)
        \label{fql}
\end{eqnarray}
where the numerator is given by, from Eq.(\ref{fqlnumap}),
\begin{eqnarray}
 {\cal N}_J^q(\vec x,L) &\equiv& <\tilde N_J^q(L)> \nonumber \\
          &=& \sum_{j_1 = (J-1)L+1}^{JL} \sum_{j_2 = (J-1)L+1}^{JL}
                  \cdots \sum_{j_q = (J-1)L+1}^{JL}
              \left[\frac{x_{j_1}}{j_1} \frac{x_{j_2}}{j_2} \cdots
                    \frac{x_{j_q}}{j_q} \right]
          \nonumber  \\  & &  \hspace{2cm} \times ~
              \frac{A!}{Q_A(\vec x)}
           \frac{Q_{A - \{j_1 + j_2 + \cdots + j_q\}}(\vec x)}
                  {(A - \{j_1 + j_2 + \cdots + j_q\})!}
                     \label{fqlnum}
\end{eqnarray}
and the denominator, from Eq.(\ref{fqldenap}), can be simplified to
\begin{eqnarray}
 {\cal D}_J^q(\vec x,L) &\equiv& <N_J(L)>^q  \nonumber \\
          &=& \left[ \sum_{j = (J-1)L+1}^{JL}
                  \left(\frac{x_j}{j}\right) \frac{A!}{Q_A(\vec x)}
                  \frac{Q_{A - j}(\vec x)}{(A - j)!} \right]^q .
                     \label{fqlden}
\end{eqnarray}
For $x=1$ in the $x$ model, since $Q_A(x=1) = A!$,
these equations give $F_q(L) = 1$
independent of $q$ and $L$ when the constraint imposed by
the $\theta$-function implicitly in Eqs.(\ref{fqlnum}) and
(\ref{fqlden}) and explicitly in Eq.(\ref{qax}) is omitted.
This result is the same as that obtained from an independent
Poisson distribution.
At $x = 1$, $<n_k> = 1/k$, a distribution which exhibits scale
invariance and self-similarity \cite{mekjian,otherx}.
When the $\theta$-function dependence is neglected,
$F_q(L)$ is very near to 1 reflecting the
Poissonian character of $P_A(\vec n, \vec x)$ of the previous subsections.
Specifically, the $F_q(L)$ is slightly less than 1 for $x > 1$
and slightly greater than 1 for $x < 1$ when this $\theta$-function
is neglected (c.f. Eq.(\ref{sigmx})).
Considering the finite size effects (effect of the $\theta$-function),
$F_q(L)$ can become quite small compared to 1 for small $x$ (see Fig.2).

The $x$ model Eq.(\ref{qax}) and
Refs.\cite{mekjian,general,otherx} allows one to evaluate
these factorial moments $F_q(L)$ exactly.
As we already noted the probability distribution for a given
cluster size, the $P_A(n_r,x)$ of Eq.(\ref{panrx}), is not
strictly Poisson even though it may be quite similar
to a Poisson distribution (see Fig.1).
For the $x$ model, from Eqs.(\ref{fql}) -- (\ref{fqlden}),
we have
\begin{eqnarray}
 F_q(L) &=& \frac{ \sum_{J=1}^{[A/(qL)]}
     \sum_{j_1 \in J} \cdots \sum_{j_q \in J}
     \left[\frac{x}{j_1} \cdots \frac{x}{j_q}\right]
       \frac{A!}{\Gamma(x + A)}
       \frac{\Gamma(x + A - \{j_1 + \cdots + j_q\})}
             {(A-\{j_1 + \cdots + j_q\})!}  }
      {   \left[\sum_{J=1}^{[A/L]}
            \left(\sum_{j \in J} \frac{x}{j} \frac{A!}{(A-j)!}
            \frac{\Gamma(x + A - j)}{\Gamma(x + A)} \right)^q \right] } .
        \label{fqlxmdl}
\end{eqnarray}
At $x = 1$, and with $A$ an integer multiple of $q L$, Eq.(\ref{fqlxmdl})
reduces to
\begin{eqnarray}
 F_q(L) = \sum_{J=1}^{[A/(qL)]}
             \left[\sum_{j=(J-1)L+1}^{JL} \frac{1}{j}\right]^q
         \left/ \sum_{J=1}^{[A/L]}
             \left[\sum_{j=(J-1)L+1}^{JL} \frac{1}{j}\right]^q \right. .
\end{eqnarray}
For the unit bin size $L = 1$, Eq.(\ref{fqlxmdl}) reduces to
\begin{eqnarray}
 F_q(L=1) &=& \frac{ \sum_{k=1}^{[A/q]} \left[\left(\frac{x}{k}\right)^q
        \frac{A!}{(A-qk)!} \frac{\Gamma(x+A-qk)}{\Gamma(x+A)} \right] }
        { \sum_{k=1}^A \left[ \left(\frac{x}{k}\right)
           \frac{A!}{(A-k)!} \frac{\Gamma(x+A-k)}{\Gamma(x+A)}\right]^{q} }
\end{eqnarray}
which can further be reduced for the $x = 1$ case to
\begin{eqnarray}
 F_q(L=1,x=1) &=& \frac{ \left[\sum_{k=1}^{[A/q]} \frac{1}{k^q}\right] }
                       { \left[\sum_{k=1}^A \frac{1}{k^q}\right]  } .
            \label{fqlx1}
\end{eqnarray}
In these equations the ranges of the sums are
different between numerator and the denominator and
this originates from the
finite $A$ effect (c.f., $\theta(A)$ in $Q_A$ of Eqs.(\ref{qax})
and (\ref{qaxy})).
Fig.2 illustrates the behavior of $F_q(L)$ as a function of $L$
for various $x$ values.
$F_q(L)$ is an increasing function of $L$ and
follows a power law behavior for small $x$ with an intermittency
exponent $\alpha(q)$ proportional to $q$ (see Table 1).
For large $x$, $F_q(L)$ decreases exponentially with a correlation
length inversely proportional to $(q-1)^2$ (see Table 1) for small $L$
and
then exhibit a saturating behavior for large $L$.
At low $x$, we can assume $<n_k>~ < 1$ which is similar to
a Fermi-Dirac distribution. Thus $\tilde n_k^q \approx 0$ for $q > 1$
and this fact gives rise to a small value of $F_q(L)$.
At $x = 1$, a behavior based on a Poisson distribution follows since
$<\tilde n_k^q> = <n_k>^q$.
At large $x$, $n_k = 0$ for $k$ larger than a fraction of $A$.
This results in the saturation of $F_q(L)$ at large $L$.
The saturation occurs at smaller $L$ for larger $x$.
In comparison, $<n_k>$ has a U shape at $x < 1$, a power law behavior
at $x \approx 1$, and an exponential behavior at large $x$.
At large $x$, only the first bin contributes to the factorial moments.
Due to finite $A$ effects, for any value of $x$, any bin $J$ constituted
with clusters of $k \ge A/2$ has the total multiplicity $N_J \le 1$
and thus does not contribute to $F_q(L)$.
At low $x$, $F_q(L)$ is quite small (far from a Poissonian behavior).
At large $x$, $F_q(L)$ is near 1, reflecting the fact that the distribution
is approximately a Poisson.

Also shown in Fig. 2 are
\begin{eqnarray}
 \tilde F_q(L) = \frac{\sum_{J=1}^{[A/L]} <\tilde N_J^q>}
                      {\left(\sum_{J=1}^{[A/L]} <N_J>\right)^q} .
\label{fqltld}
\end{eqnarray}
This quantity is normalized with total multiplicity $<M> = \sum_J <N_J>$
rather than the average multiplicity $<N_J>$ of each bin.
For a uniform distribution, $\tilde F_q(L) \propto (L/A)^{(q-1)}$
and $\tilde F_q(L)$ becomes 1 for $n_k$ peaked in one bin.
This quantity shows a power law behavior with positive intermittency
power $\beta(q)$
proportional to $q$ (see Fig.2c and Table 1).

Situations in which events with different values of $x$ are mixed can also
be considered.
Since the corresponding probability of a partition $\vec n$ is
a linear combination of $P_A(\vec n, \vec x)$ over all the values of $x$,
\begin{eqnarray}
 P_A(\vec n) \equiv \sum_{\vec x} a(\vec x) P_A(\vec n, \vec x) ,
\label{pamix}
\end{eqnarray}
any factorial moments $<\tilde N_J^q>$ is also a linear combination
of $<\tilde N_J^q>$ for each $x$. Thus we have
\begin{eqnarray}
 F_q(L) &=& \frac{\sum_{J=1}^{[A/L]} \left[\sum_x a(x) <\tilde
N_J^q(x)>\right]}
            {\sum_{J=1}^{[A/L]} \left[\sum_x a(x) <N_J(x)>\right]^q}
         \nonumber \\
     &=& \sum_{J=1}^{[A/L]} \sum_x
         \left[\frac{a(x) <N_J(x)>^q}{\sum_{J=1}^{[a/L]} <M_J>^q}\right]
         \left[\frac{<\tilde N_J^q(x)>}{<N_J(x)>^q}\right] ,  \label{fqlxs}
\end{eqnarray}
where the weight $a(x)$ is normalized as $\sum_x a(x) = 1$
and the total mean multiplicity $<M_J>$ in the bin $J$ is
$<M_J> = \sum_x a(x) <N_J(x)>$.
As can be seen from Eq.(\ref{fqlxs}), the numerator depends on $a(x)$
linearly while the denominator varies as $a(x)^q$.
Thus even though $<\tilde N_J^q(x)>/<N_J(x)>^q \approx 1$ for each bin
and for each $x$,
$F_q(L)$ can be quite large if $a(x)$ is small for the $x$ having
largest contribution of $\left[a(x) <N_J(x)>\right]$.
This type of mixing of different $x$'s has been considered in
Ref.\cite{xmix}.
The resulting values of $F_q(L)$ can become large as discussed
in Refs.\cite{ploszaj,fracnuc2}.

Some other interesting features of the evaluation of $F_q(L)$
using the $x$ model are as follows.
Defining
\begin{eqnarray}
 {\cal N}_q(L) &\equiv& \sum_{J=1}^{[A/L]} {\cal N}_J^q(x,L)
      = \sum_{J=1}^{[A/L]} <\tilde N_J^q(L)>
             \nonumber \\
 {\cal D}_q(L) &\equiv& \sum_{J=1}^{[A/L]} {\cal D}_J^q(x,L)
      = \sum_{J=1}^{[A/L]} <N_J(L)>^q   \label{ndql}
\end{eqnarray}
the factorial moment is given by $F_q(L) = {\cal N}_q(L)/{\cal D}_q(L)$.
The sum of $J$ in ${\cal N}_q(L)$ actually runs only up to
$\tilde J = [A/qL]$ since ${\cal N}_J^q(L)$ of Eq.(\ref{fqlnum}) is
zero for $J > [A/qL]$
due to the $\theta$-function appearing
in the partition function $Q_A(x)$ of Eq.(\ref{qax}).
At $x = 1$, the correlations and fluctuations are such that
the expectation value of the factorial moment in the $J$'th bin
is just the $q$'th power of the multiplicity in the $J$'th bin.
Thus ${\cal N}_J^q(x,L) = {\cal D}_J^q(x,L)$
(c.f., Eqs.(\ref{fqlnum}) and (\ref{fqlden}))
except the numerator ${\cal N}_q(L)$ is summed only
up to $[A/(qL)]$ compared to the denominator ${\cal D}_q(L)$
which is summed up to $[A/L]$ (c.f. Eq.(\ref{fqlx1})).
At $x = 1$ in the $x$ model, the mean cluster distribution
$<n_k>$ exhibits scale invariance and self-similarity
\cite{mekjian,otherx}.

An approximate form for the factorial moments for large $L$ and
$A$ at $x = 1$ can also be obtained as follows.
At $x = 1$, $<n_k> = 1/k$ and
\begin{eqnarray}
 <N_1> &=& \sum_{k=1}^L \frac{1}{k}
       \approx \log L + \gamma = \log \beta L ,  \nonumber \\
 <N_J> &=& \sum_{k=1}^{JL} \frac{1}{k} - \sum_{k=1}^{(J-1)L} \frac{1}{k}
       \approx \log (J/(J-1))  \hspace{2em} {\rm for}
                    \hspace{2em} J \ne 1 .
\end{eqnarray}
Here $\gamma = \log \beta = 0.57721\cdots$ is Euler's number.
Using these results, the $F_q(L)$ can be obtained from
\begin{eqnarray}
 F_q(L) = \frac{{\cal N}_q(L)}{{\cal D}_q(L)}
\end{eqnarray}
with ${\cal N}_q(L)$ and ${\cal D}_q(L)$ given by Eq.(\ref{ndql}).
Further approximation give, for the $x = 1$ case,
\begin{eqnarray}
 F_q(L) \approx 1
    - \frac{\sum_{J = A/(qL)}^{A/L -1} (1/J)^q}
           {(\log \beta L)^q + \sum_{J=1}^{A/L-1} (1/J)^q}
\end{eqnarray}
using $\log (1 + 1/J) \approx 1/J$.
Approximating the sum in the denominator as a zeta function
$\zeta(q)$ and approximating the numerator as $(A/L) - (A/qL)$ terms
of the order of $(qL/A)^q$ results in
\begin{eqnarray}
 F_q(L) \approx 1
     - \frac{\left(\frac{L}{A}\right)^{q-1} q^{q-1} (q-1)}
            {(\log \beta L)^q + \zeta(q)}
    \approx
     \exp\left[ - \frac{\left(\frac{L}{A}\right)^{q-1} q^{q-1} (q-1)}
            {(\log \beta L)^q + \zeta(q)} \right]
\end{eqnarray}
for $qL < A$. This result for the scaled factorial moment
exhibits the existence of a correlation length at $x = 1$
rather than a power law behavior.
In Fig.2, $F_q(L)$ stays very near one for $L < A/2$,
then $F_q(L)$ decreases exponentially for larger $L$.
For $x \approx 1$, the above result can be shown to be
\begin{eqnarray}
 F_q(L) \approx 1
    - \frac{(2-x)\left(\frac{L}{A}\right)^{q-1} q^{q-1} (q-1)}
           {(\log\beta L)^q + \zeta(q) - q(x-1) \left(\frac{L}{A}\right)
                \left[(\log\beta L)^{q-1} + \zeta(q-1)\right]}
\end{eqnarray}

\subsection{Cumulative Mass Distribution and its Staircase Properties}

\hspace{30pt}
This subsection considers
a probability
distribution associated with the mass distribution
using
\begin{eqnarray}
 P_A(k,\vec x) = \frac{k}{A} n_k = \frac{m_k}{A}  \label{pakx}
\end{eqnarray}
where $m_k$ is the total mass in the clusters of size $k$.
The fraction of mass in clusters of size $k$ can be used
to define a probability function \cite{mekjian,otherx}
given by Eq.(\ref{pakx})
since
\begin{eqnarray}
 && P_A(k,\vec x) \ge 0 , \nonumber \\
 && \sum_{k=1}^A P_A(k,\vec x) = 1 .
\end{eqnarray}
The total mass $A$ is strictly conserved in our canonical ensemble
approach.
The properties of the canonical ensemble average $<P_A(k,\vec x)>$
are studied in Ref.\cite{otherx}.
The $P_A(k, \vec x)$ is a probability distribution
for each partition $\vec n$ or event;
each event gives the distribution of nucleons in a one-dimensional
space given by the cluster size $k$.
Specifically, each point in the ensemble of partitions in $\vec n$ space has
one distribution $P_A(k, \vec x)$ corresponding to that event $\vec n$.

The ensemble averaged quantity
$<P_A(k,\vec x)> = k <n_k>/A$ is a Polya-Eggenberger
distribution for the $x$ model \cite{otherx} which is given by
\begin{eqnarray}
 <P_A(k,x)>
    = x \frac{(A-1)!}{(A-k)!} \frac{\Gamma(x+A-k)}{\Gamma(x+A)} .
\end{eqnarray}
The $<P_A(k,x)>$ can be generated from a Polya urn
with replacement \cite{otherx}. Specifically, a urn consists of $R$ red
balls and $G$ green balls.
Each time a ball is drawn, that ball and $S$ additional balls
of the same color are replaced.
The probability of drawing $m$ red balls in $n$ trials from
this urn is the Polya distribution.
The $x$ model $<P_A(k,x)>$ follows when $m = k-1$,
$n = A-1$, $R = S$, and $x = G/S$ as discussed in Ref.\cite{otherx}.

At $x = 1$ in the $x$ model, the mass $A$ is uniformly distributed
over the clusters so that $<m_k> = k <n_k> = 1$, independent of $k$.
At $x \ne 1$, the mass in clusters of size $k$ becomes an increasing
function with increasing $k$ for $x < 1$
and a decreasing function of $k$ for $x > 1$.

If we neglect the discreteness of the size index $k$,
then at $x = 1$, the mass distribution is analogous to
that of a uniform bar where $k$ represents the distance
along the bar and the length of the bar is $A$.
By treating each nucleon as a uniform bar of unit mass with
a unit length in cluster size space, $P_A(k,\vec x)$
becomes a local probability density which is uniform in
the continuous variable $s$ in the range of $k - 1 \le s < k$.
Then the mass density $m(s) = m_k$ is constant in this range.
A cumulative mass defined by
\begin{eqnarray}
 M(k,\vec x) = \int_0^k m(s) d s = A \int_0^k P_A(s,\vec x) d s
        \label{cummass}
\end{eqnarray}
forms a staircase with a sloped rise at each step
in contrast to the staircase of a cumulative mass considered
in Refs.\cite{mekjian,otherx} where each step is flat
and has a vertical rise.
Plotting this cumulative mass as shown in Fig.3, the total height
and the width are fixed to $A$ for any partition $\vec n$.
All fragmentation staircases start from the point
$(k,M(k)) = (0,0)$ and end at the point $(k,M(k)) = (A,A)$.
Projecting a staircase onto the vertical axis
gives a discrete set of intercept
points analogous to an energy level diagram (see Fig.3).
With this interpretation,
the mass $m_k$ corresponds to a ``level spacing''.
The height at each step is the relevant quantity.
Degeneracy appears in this staircase diagram as a
flat stair with zero height and non-degeneracy appears
as a rise with non-zero height.
In Section 4 we will consider the staircase behavior of the data of
Ref.\cite{nucdat} and compare this data with the predictions of the $x$-model.

\subsection{Power Moments of a Probability Distribution}

\hspace{30pt}
Our study of factorial moments in Subsection 2.2
were based on a probability distribution $P_A(\vec n, \vec x)$
in the $A$ dimensional multiplicity space $\vec n$;
ensemble of points in $\vec n$ space.
In this subsection, we will consider power moments of the
mass distribution and probability measures associated with them.
In Eq.(\ref{pakx}) in Section 2.3, we introduced the probability
distribution $P_A(j,\vec x)$ related to the mass in a cluster
of size $j$.
We now investigate the scaling properties associated with these
probability measures. Previous concerns have focussed on factorial
moments \cite{ploszaj,fracnuc1,fracnuc2}
and here we will present a different approach based on
the distribution of mass in a fragmentation process.
For this purpose we introduce following quantity:
\begin{eqnarray}
 {\cal P}_q(L) = \sum_{J=1}^{[A/L]} \left< \left[\sum_{j=(J-1)L+1}^{JL}
                    P_A(j, \vec x) \right]^q \right> .  \label{pqldef}
\end{eqnarray}
For a unit bin size $L = 1$, this reduces to
\begin{eqnarray}
 {\cal P}_q(1) = \sum_{k=1}^{A} < P_A^q(k, \vec x) >
\end{eqnarray}
which is $q$'th moment of $P_A(k, \vec x)$.
For $q = 2$, Eq.(\ref{pqldef}) reduces to
\begin{eqnarray}
 {\cal P}_2(L) = \sum_{k=1}^A <P_A^2(k, \vec x)>
     + \sum_{J=1}^{[A/L]} \sum_{j \in J} \sum_{k\ne j \in J}
         <P_A(j, \vec x) P_A(k, \vec x)>
\end{eqnarray}
which is a sum of the second order power moment of $P_A(k, \vec x)$
and the correlations between different size clusters $k$ and $j$
up to the distance $L-1$.
Thus, departures of the ratio ${\cal P}_2(L)/{\cal P}_2(1)$ from unity
quantifies the correlations in the mass distribution
between different size clusters.
If the distribution has an underlying scale or coherence length $\xi$,
then this ratio would have a structure of the form
\begin{eqnarray}
 \frac{{\cal P}_q(L+1) - {\cal P}_q(L)}{{\cal P}_q(1)} \propto e^{-L/\xi}
\end{eqnarray}
On the other hand, if the distribution is scale invariant then
the ratio has a power law behavior,
\begin{eqnarray}
 \frac{{\cal P}_q(L)}{{\cal P}_q(1)} \propto L^{-\beta}
     = \frac{1}{\Gamma(\beta)} \int_0^\infty \frac{d \xi}{\xi^{\beta+1}}
            e^{- L/\xi} ,
\end{eqnarray}
with a critical exponent $\beta$.
As noted in Ref.\cite{otherx}, a power law behavior is associated with
a
distribution of length scales.

Fig.4 shows the power moment ${\cal P}_q(L)$ in the $x$ model described
in section 2. Specifically, from Fig. 4 we see that the power
moment exhibit
a power law behavior for large $x$ and an exponential behavior
for small $x$.
At small $x$, which corresponds to a low temperature,
the mean interaction between nucleons
is dominant compared to the thermal energy
resulting in the formation of large clusters.
Thus one size of cluster is correlated with
other clusters in the system event by event.
On the other hand, at large $x$, which corresponds to high
temperature, thermal energy becomes dominant and mean field
effect can be neglected resulting in the appearance of many nucleons and
small clusters.
As a result, one cluster is independent
of the other clusters except for constraints imposed by
total number conservation.
Also, at large $x$ large clusters have a very small probability
of being present. These effects results in a scale invariance
of the distribution.
Table 2 shows the intermittency exponent and the correlation
length for various $x$. The
results obtained exhibit a power law behavior with the intermittency
exponent proportional to $(q-1)$ for large $x$.
For small $x$, they show an exponential increase with a
correlation length
inversely proportional to $q$.
For intermediate values of $x$, the power moment shows neither
the power law behavior nor the exponential behavior.

We can also define a scaled power moment as
\begin{eqnarray}
 {\cal P}_q^s(L) \equiv
 \frac{{\cal P}_q(L)}{\sum_{J=1}^{[A/L]} \left[\sum_{j=(J-1)L+1}^{JL}
                      <P_A(j, \vec x)> \right]^q}
      \propto \left(\frac{L}{A}\right)^{\alpha_q}   \label{sclpql}
\end{eqnarray}
which is similar to the scaled factorial moment of Subsection 2.2.
This scaled power moment can be reduced to a scaled factorial moment
of the mass distribution using the relations
between power and factorial.
The size $L$ dependence of the denominator of Eq.(\ref{sclpql}) is
shown in Fig.4c and Table 3.
The scaled power moment of Eq.(\ref{sclpql}) measures the departure of
the distribution from its mean behavior in a manner similar
to the scaled factorial moment.
If ${\cal P}_q^s(L)$ is proportional to a power of the bin size $L$,
i.e., $L^{\alpha_q}$, the power $\alpha_q$ is
called an intermittency exponent \cite{ploszaj,fracnuc1,fracnuc2}.
If $P_A(k, \vec x)$ were uniform over $k$
with $P_A(k, \vec x) = <P_A(k, \vec x)> = 1/A$,
then the denominator in the scaled power moment of Eq.(\ref{sclpql})
is simply $(L/A)^{q-1}$;
\begin{eqnarray}
 {\cal P}_q(L) \approx \left(\frac{L}{A}\right)^{q-1} {\cal P}_q^s(L) .
     \label{psprel}
\end{eqnarray}
Due to this denominator, the power moment ${\cal P}_q(L)$ can be
an increasing function of $L$ even though the scaled power moment
decreases as $L$ increases.
However, if all the mass is in one cluster size, say $k$,
then both the power moment ${\cal P}_q(L)$ and the scaled
power moment ${\cal P}_q^s(L)$ are one.
These results can be summarized as
\begin{center}
\begin{tabular}{c|c|c}
   \hline  & \\
 $P_A(k, \vec x)$  &   uniform    &   peak in one $k$     \\
               & ($1/A$ for all $k$)   & ($P_A(k,x) = 1$ and zero otherwise)
\\
   \hline  &  \\
 ${\cal P}_q(L)$   & $(L/A)^{q-1}$    &   1   \\  &  \\
 ${\cal P}_q^s(L)$ &           1      &   1   \\  &  \\
   \hline
\end{tabular}
\end{center}
Both these special cases, the uniform case and the case of
a peak in one $k$, have no fluctuation or correlation.
On the other hand, $P_A(k,\vec x)$ itself follows a power law
($L/A$ in a bin of legnth $L$) for the uniform case
as the size of the system $A$ changes
while it stays unchanged for the case of one $k$ peak.
This shows that the size $L$ dependence of the scaled power
moment ${\cal P}_q^s(L)$ characterizes a bin size dependence
of the fluctuation or the
correlation compared to an independent Poisson distribution
or to a mean field approximation of the distribution. This feature
should be contrasted with
the power moment ${\cal P}_q(L)$ which
characterizes the size dependence or scaling properties
of the distribution itself.
A similar relation also holds between the scaled factorial moment
and the factorial moment also.
Thus, power or factorial moments characterize better
the scaling property of a distribution.
On the other hand, the fluctuation and
correlations of a distribution compared to a mean field approximation
are better studied using the scaled power or scaled factorial moment.
Fig.4(c) also shows the power moment of the mean mass distribution
$<P_A(k,x)>$. This result shows a power law behavior for
any value of $x$. The power is small for a small $x$ and the power
stays at about $(q-1)$ for $x \ge 1$ until $x$ becomes quite large.
This power is essentially the power appearing in the above table.
At large $x$ the power moment saturate to 1 at small $L$.

\section{Power Moments and Renyi Entropies;
Dimensions and Intermittency}

\hspace*{30pt}
In previous section, we have studied various type of probability
distributions related to nuclear fragmentation
and the bin size (scale) dependence of the factorial and power moments.
We have also considered the fluctuations and correlations
related to these probability distributions.
This section discusses the concept of assigning a generalized
entropy and a measure or dimension to
various distributions developed in the previous section.
The question of a true measure assigned to the discrete
system (the cluster index $k$ is discrete)
becomes somewhat problematic because a point has a zero
measure mathematically.
It should be noted that one can discuss properties of the
distribution of fragments without introducing ideas related
to measure or dimension.
We could just as well discuss correlations, fluctuations,
Nevertheless, a discussion of a generalized dimension
based on a measure will be given because of the recent
discussion and interest in it.
Our discussion will focus more on dimensions associated with
probability distributions of the previous section
and on associated Renyi entropies.
These probability measures are obtained from the clusters of size $k$
and from the mass in bins of length $L$.
The exact canonical ensemble developed in Ref.\cite{mekjian,general}
and applied here allows us to consider properties associated
with these distributions.
The approach of others is based more on factorial moments associated
with the cluster numbers $n_k$.
Hopefully, our approaches will compliment and widen the factorial
moment approach.

\subsection{Generalized Renyi Entropies and Generalized Dimension;
Self-Similar Substructure}

\hspace*{30pt}
The concept of the covering dimension \cite{fractal} can be
extended by considering a generalized Renyi entropy and
an associated generalized dimension.
The generalized dimension given below includes not only the
Hausdorff covering dimension but also the information dimension
associated with the Shannon information content of a probability
distribution and the correlation dimension.
The correlation dimension comes from a generalized entropy
introduced by Renyi which includes the Shannon information.
The Renyi entropies relate to correlation functions
of a probability distribution.
We discuss these generalized entropies and generalized dimensions
in the framework of
probability distributions considered in Section 2 for a
nuclear fragmentation.

When we group $A$ data points into bins of length $r = L/A$
the number of bins $N$ becomes $[A/L]$ with each bin
having $L$ primary data points except the last bin.
When $A$ is not an integer multiple of $L$, the last bin is smaller
than the other bins.
However, if we can neglect the effect of the smallness of the final bin,
or when all the bin size $r_J$'s are of equal length $r = L/A$,
we can define the generalized dimension $D_q$ for a given probability
distribution $P_J$ through a condition \cite{fractal} of
\begin{eqnarray}
 H_q = \left[\sum_{J=1}^N p_J^q \right] r^{-(q-1) D_q}
     = \left[\sum_{J=1}^N \left(\frac{p_J}{r^{D_q}}\right)^q \right]
        r^{D_q} = 1 ,   \label{hqrcnt}
\end{eqnarray}
for $-\infty < q < \infty$. The $H_q$ is called the partition sum or
total measure.
A simple quantity defined by
\begin{eqnarray}
 S_q(L) = \frac{-1}{q-1} \log \left[\sum_{J=1}^N p_J^q\right]
         \ge 0    \label{sqgen}
\end{eqnarray}
is useful.
The $S_q(L)$ is called the generalized Renyi entropy
of order $q$ associated with the probability distribution
$p_J$ for $N = A/L$ data points.
For equal bins of length $r = L/A$ each bin has $L$ data points.
The generalized information dimension is now simply given by
\begin{eqnarray}
 D_q(L) = \frac{1}{q-1} \frac{\log\left[\sum_{J=1}^N p_J^q\right]}{\log r}
     = - ~\frac{S_q(L)}{\log r}
   \hspace{2em} {\rm for} \hspace{2em} -\infty < q < \infty  \label{dqgen}
\end{eqnarray}
with $ q = 1$ taken as a limit.
Eq.(\ref{hqrcnt}) relates the information dimension with the power
moment considered in Subsection 2.4.
Here the generalized information dimension $D_q(L)$
depends on the bin size $L$ or the number of bins $N$
and it becomes the generalized fractal dimension \cite{fractal}
in the limit of $r = L/A \to 0$ or $N = A/L \to \infty$.

As $q \to 0$, $\left[\sum_J \left(p_J/r^{D_q}\right)^q\right] = N_q$
becomes the total number of boxes $N_0$ of size $L$
covering the region having non-zero probability $p_J$.
Here $N_0 \le N$ with the equality occurring when all the bins
have non-zero probability.
In the limit $q \to 1$, $S_1(L)$ is
\begin{eqnarray}
 S_1(L) &=& - \sum_{J=1}^N p_J \log p_J .
\end{eqnarray}
which is recognized as the Shannon information.
The associated generalized information dimension $D_1(L)$ is
\begin{eqnarray}
 D_1(L) &=& \frac{\sum_{J=1}^N p_J \log p_J}{\log r}
         =  \frac{- S_1}{\log r} .
\end{eqnarray}
The fractal information dimension is obtained from $D_1(L)$
as $r = L/A \to 0$.

Eq.(\ref{hqrcnt}) relates the power moment considered in Subsection 2.4
with the generalized information dimension $D_q$
and the generalized entropy $S_q$.
If the data set has $N'$ bins with the same probability
$p_J = 1/N'$ and if the remaining bins ($N - N'$) have $p_J = 0$,
then the dimension $D_q = \log N' / \log N$ is independent
of the order $q$ (c.f., monofractal \cite{fractal} or single process).
Thus, the $q$ dependence of $D_q$ means the data set has
non-zero probabilities $p_J$ which depend on the bin $J$
(c.f., multifractal \cite{fractal} or multiprocess).
For $q \to \infty$, only the bins with the highest probability
$p_{max}$ count, $D_\infty = \log N_{max} / \log N$,
where $N_{max}$ is the number of bins with $p_{max}$ and $r = 1/N$.
For $q \to -\infty$, only the bins with the nonzero lowest probability
$p_{min}$ count, $D_{-\infty} = \log N_{min} / \log N$.
On the other hand, if the data set has a constant probability $p_J = p$
with varying bin size $r_J$, then we have $D_\infty = \log p / \log r_{min}$
and $D_{-\infty} = \log p / \log r_{max}$.
For a scale invariant system, such as a self-similar Cantor set,
the generalized dimensions $D_q$
are independent of bin size $L$.
For strictly self-similar mathematical monofractals
(all the nonzero $p_J$ are the same value $p$), such as the
Sierpinski gasket or the Cantor dust, all the fractal dimensions
$D_q(L \to 0)$ are $q$ independent \cite{fractal}.

For $q = 2$, $S_2$ contains $p_J^2$ and corresponds to the
probability that two elements belong to the same cell $J$.
Thus $p_J^2$ measures correlations and fluctuations.
For unit bin size, only one cluster size
belongs to each bin and $r = L/A = 1/A$.
Using the mass distribution $P_A(k, \vec x) = k n_k / A$ of
Eq.(\ref{pakx}), $S_2(L)$ for $L = 1$ is
\begin{eqnarray}
 S_2(1) = - \log \left[ \sum_{k=1}^A \frac{k^2 n_k^2}{A^2} \right] .
\end{eqnarray}
The $S_2$ sum rule with $L = 1$
determines the generalized dimension $D_2(1) = - S_2(1) / \log (1/A)$
for $A$ data points with the bin size $r = 1/A$.
When $L = 2$, two clusters belong to each bin (except for the last
cell for odd $A$).
The $S_2(L)$ for $L = 2$ is given by
\begin{eqnarray}
 S_2(2) &=& - \log \left(\sum_{J=1}^{[A/2]} \left[P_A(2J-1, \vec x)
                     + P_A(2J, \vec x)\right]^2 \right)   \nonumber \\
     &=& - \log \left[ \sum_{k=1}^A \frac{k^2 n_k^2}{A^2}
        + \sum_{k=1,3,5,...} \frac{k(k+1) n_k n_{k+1}}{A^2} \right] .
\end{eqnarray}
$S_2(2)$ contains not only $n_k^2$ but also
correlations from the cross terms $n_k n_{k+1}$.
Larger cell sizes (larger $L > 2$) contain more correlation
terms $n_k n_{k+j}$, with $j = 0$, 1, ..., $(L-1)$.
Thus $S_2(L)$ is called the correlation entropy
and $D_2(L)$ is the associated correlation dimension of bin size $L$.
Higher $q$ contains higher order correlations
$n_{k_1} n_{k_2} \cdots n_{k_q}$.

 From Eq.(\ref{hqrcnt}), we can define a generalized cumulative
measure (mass) as
\begin{eqnarray}
 H_q(K) = \sum_{J=1}^K ~ p_J^q ~ r_J^{(1-q) D_q}
     = \sum_{J=1}^K \left[\frac{p_J}{r_J^{D_q}}\right]^q r_J^{D_q}
     = \int_0^K \left[\frac{p(r)}{r^{D_q}}\right]^q
        d \left[ r^{D_q} \right] .
                \label{hqcum}
\end{eqnarray}
The $H_q(K)$ of Eq.(\ref{hqcum}) forms a staircase with each
step height of $\left[p_J/r_J^{D_q}\right]^q r_J^{D_q}$
and step width of $r_J^{D_q}$.
Treating each bin such that the probability
density $p_J/r_J^{D_q}$ is taken as constant in a bin,
we can define the integral representation given in Eq.(\ref{hqcum}).
This procedure gives a sloped staircase as discussed in subsection 2.3.
At the step where $p_J^q = 0$ the staircase is flat (horizontal), and
is similar to a void in the Cantor set \cite{fractal}.
$H_q = H_q(N)$ is the total height of the staircase and $D_q$ is the
dimension which is required to set $H_q = 1$.
For the case of $r_J = r \to 0$,
for $q = 0$, $d\left[r^{D_0}\right]$ is the measure of a fractal
space \cite{barnsley}.

\subsection{Ensemble Average}

\hspace*{30pt}
In previous subsection, we considered a probability distribution
$p_J$ in a space made of bins for a partition $\vec n$ or an event.
Here the measure $H_q$ or the power moment $\sum_J p_J^q = e^{-(q-1) S_q(L)}$
corresponds to an average of $p_J^{(q-1)}$ over the bins for a single event
(horizontal average).
For an ensemble of many events, we can also look at
the generalized dimension and entropy of the ensemble.
However, for this case, ensemble averages over
many events (vertical average) within a bin
besides the horizontal average should be taken.
There are two simple way to consider these two different averaging
processes.

One way is first simply taking the ensemble average $\left<p_J\right>$
in a bin $J$, then taking $q$'th power of this $\left<p_J\right>$
(i.e., a horizontal average of $<p_J>^{(q-1)}$)
to calculate the measure $H_q$ or the entropy $S_q(L)$ for the ensemble.
This procedure gives exactly the same relations as $H_q$ itself for
a one event system as discussed in the previous subsection. Also such
a procedure corresponds
to a mean field approximation in statistical physics  and misses
fluctuation effects.
As an example, the generalized Renyi entropy and the generalized
dimension are given by
\begin{eqnarray}
 S_q(L) &=& \frac{-1}{q-1} \log \left[\sum_{J=1}^N <p_J>^q\right]
        \hspace{1cm} {\rm for} \hspace{1cm} q \ne 1 ,  \nonumber \\
 S_1(L) &=& - \sum_{J=1}^N <p_J> \log <p_J> ,   \nonumber\\
 D_q(L) &=& \frac{- S_q(L)}{\log(L/A)} .   \label{sqmeanp}
\end{eqnarray}
The whole ensemble is then treated as an event with the mean
mass distribution $<p_J> = <P_A(k, \vec x)> = k <n_k> / A$
and $r_k = 1/A$ for all $k$.
The $S_q(L)$ is related to the denominator of the scaled power
moment given by Eq.(\ref{sclpql}).
As an example, at $x = 1$ in our $x$ model, we have $<P_A(k,x)> = 1/A$
\cite{mekjian,general,otherx}
which gives  $<p_J> = L/A$ and $r = L/A$ for a bin size of $L$.
Then the $H_q$ of (\ref{hqrcnt}) is
\begin{eqnarray}
 H_q = \sum_{J=1}^{[A/L]} \left(\frac{L}{A}\right)^q
       \left(\frac{L}{A}\right)^{(1-q)D_q(L)}
  = \left(\frac{L}{A}\right)^{(q-1)(1 - D_q(L))} .
\end{eqnarray}
Consequently the condition of $H_q = 1$ gives $D_q(L) = 1$ for
any bin size $L$ and for any order $q$.
This generalized dimension $D_q(L)$ is the same as the Euclidean
topological dimension (i.e., no holes or voids)
and exhibits a scale invariance ($L$ independence)
reflecting the self-similar substructure.
The $q$ independence of $D_q(L)$ represent a monofractal
structure (single process) of this case.
Using this self-similar property of this model, we can treat
the $x = 1$ case as a fractal system having an infinite
self-similar substructure with $r = L/A \to 0$ or $A \to \infty$.
Then $D_q(L)$ stays constant as $r = L/A \to 0$ and
becomes the generalized fractal dimension.
In Ref.\cite{otherx}, we discussed the scale invariance existing
in the $x = 1$ case using the cluster size distribution $<n_k>$.
Fig.4(c) shows the power moments of $<p_J>$ in the $x$ model
(see Subsection 2.4). This figure indicates that the generalized
dimension $D_q(L)$ (see Table 3 also) is much smaller than 1 for small $x$
and stays near to 1 for large $x$, i.e., at high temperature.
At much larger $x$, $D_q(L)$ decreases again with a $q$ dependence.

In the second way of ensemble averaging,
we first find the measure $H_q$  event by event,
then take an ensemble average of this $H_q$ of each event.
In the equal bin size case ($r_J = r = L/A$),
this averaging process gives
\begin{eqnarray}
 \left< H_q \right>
   = \sum_{J=1}^N \left< \left[\frac{p_J}{r_J^{D_q}}\right]^q\right> r_J^{D_q}
   = \sum_{J=1}^N \left< p_J^q\right> ~ r^{-(q-1) D_q} .
\end{eqnarray}
Using Eq.(\ref{hqrcnt}), the condition of $\left<H_q\right> = 1$
determines the generalized
dimension $D_q(L)$ of the ensemble and the generalized Renyi
entropy for the ensemble;
\begin{eqnarray}
 S_q(L) &=& \frac{-1}{q-1} \log \left[\sum_{J=1}^N <p_J^q>\right]
        \hspace{1cm} {\rm for} \hspace{1cm} q \ne 1 ,  \nonumber \\
 S_1(L) &=& - \sum_{J=1}^N <p_J \log p_J> ,   \nonumber  \\
 D_q(L) &=& \frac{-S_q(L)}{\log(L/A)} .   \label{sqmeanh}
\end{eqnarray}
This ensemble averaged entropy $S_q(L)$ represents
the $q$-th order moment of $p_J$ over the ensemble in each cell.
As a special example, for $q = 2$,
\begin{eqnarray}
 S_2(L) &=& - \log \left(\sum_{J=1}^{[A/L]}
                 \left<\left[\sum_{j\in J} p_j \right]^2\right> \right)
                         \nonumber \\
     &=& - \log \left[ \sum_{k=1}^A \frac{k^2 <n_k^2>}{A^2}
      + \sum_{j \in J} \sum_{k \ne j \in J} \frac{j k <n_j n_k>}{A^2} \right].
\end{eqnarray}
Thus $S_2(L)$ contains not only $<n_k^2>$ but also
correlations from the cross terms $<n_k n_{j}>$
with $|k-j| = 1$, 2, ..., $(L-1)$.
Higher $q$ contains higher order correlations
$\left<n_{k_1} n_{k_2} \cdots n_{k_q}\right>$.
If, instead of the power moment $\sum_J <n_J^q>$,
the scaled moment defined as $\sum_J <n_J^q>/\sum_J <n_J>^q$
is used as in Ref.\cite{fracnuc1,fracnuc2} or in subsection 2.2,
then
\begin{eqnarray}
 S_1(L) &=& \frac{\sum_J <n_J \log n_J>}{\sum_J <n_J>}
           - \frac{\sum_J <n_J> \log <n_J>}{\sum_J <n_J>}   \nonumber \\
        &=& \sum_{J=1}^N \left<\left(\frac{n_J}{\sum_J <n_J>}\right)
               \log \left(\frac{n_J}{\sum_J<n_J>}\right) \right>  \nonumber \\
  & & \hspace{1.5cm} - \sum_{J=1}^N \left(\frac{<n_J>}{\sum_J <n_J>}\right)
               \log \left(\frac{<n_J>}{\sum_J <n_J>}\right) .  \label{sqmn1}
\end{eqnarray}
For the $n_J = p_J$ case, $\sum_J n_J = \sum_J p_J = 1$.
The difference between power moment and the scaled power moment for
$S_1(L)$ can be seen by comparing Eq.(\ref{sqmeanh}) and Eq.(\ref{sqmn1}).

Fig.4 shows the behavior of the power moment
${\cal P}_q(L) = \sum_J <p_J^q>$
and the corresponding intermittency power and the correlation length
are shown in Table 2.
This figure illustrates the exponential behavior at small $x$
and the power law behavior (scale invariance) at large $x$.
At the intermediate $x$, this figure shows neither the power law
nor the exponential behavior.
The corresponding dimension $D_q(L)$ is 0.75 for $x = 20$
and 0.83 for $x = 50$ independent of $q$ (monofractal behavior).
At $x = 79$, $D_q(L)$ is 0.80 and 0.76 for $q = 2$ and 3
respectively exhibiting the $q$ dependence, similar to the small $x$ case.

\section{Analysis of Data in Nuclear Fragmentation}

\hspace*{30pt}
In this section we will analyze the emulsion data of Ref.\cite{nucdat}.
This data has already been discussed by other groups \cite{ploszaj,fracnuc2}
using a scaled factorial moment approach.
Our approach to the analysis of this data
will be different than that given by these other groups. In particular
we will study this data using the probability distributions associated
with the fraction of the mass in cluster of various sizes, and various
powers of this quantity which give the generalized Renyi entropies.
Also the staircase properties of the cumulative mass distribution
associated with this data are studied.

The data of Ref.\cite{nucdat} was obtained from an emulsion experiments for
$^{197}$Au$^{79}$ at 0.99 GeV/amu and consists of 415 events, all conserving
the total charge. These 415
events are a very small subset of the total number of possible partitions
of the Au into all its possible fragmentation modes. The fragments are
identified by their charges so that the initial $Z=79$ is partitioned
into ($n_1, n_2,...,n_{79}$), where $n_z$ equals the number
of fragments of charge $z$ in a given event. The probability
distributions will be given in terms of the charges,
$P_Z(z) = z n_z/Z$, with $z = 1$, 2, ..., 79. The cumulative charge
distribution, the analog of Eq.(\ref{cummass}), is
\begin{equation}
  M(s) ~=~ \sum_{z=1}^s ~ z n_z .   \label{aucumas}
\end{equation}
Ensemble average quantities can be obtained from the 415 events
straightforwardly. The results of our approach will now be given.

The first requirement of a model is to reproduce the inclusive data.
Fig.6 shows a fit to the charge distribution $<n_z>$ using the
$x$-model. Apart from the large even-odd oscillation, the $x$-model
predicts the U-shaped feature of the data. The value of $x$ used was
$x$ = 0.3 (dashed curve).
Some discrepancy between the theoretical fit and the
experimental data appears at low $z$, where the theory underestimates
the experimental data for this value of $x$. The data suggests a
higher value of $x$ for small $z$.
In the single $x$ model, the distribution $<n_z>$ exhibits a U shape at a
small $x < 1$ and an exponentially fast drop at a large $x > 1$ (see
figures in Refs.\cite{xymodel,nucfit}).
Also shown by a dotted line is a linear combination of $x_1 = 0.5$
and $x_2 = 50$ (c.f., Eqs.(\ref{pamix}) and (\ref{fqlxs})),
\begin{eqnarray}
 P_A(\vec n, x_1, x_2) = a(x_1) P_A(\vec n, x_1)
        + a(x_2) P_A(\vec n, x_2) ,   \label{ptwox}
\end{eqnarray}
with $a(x_1) = 0.66$ and $a(x_2) = 0.34$.
The $x = x_2 = 50$ component fits the fast drop at the low charge region
and the $x = x_1 = 0.5$ component fits the U shape behavior at
the higher $z$ part.

Fig.7a shows the cumulative charge distribution of Eq.(\ref{aucumas}).
For $x$=0.3, the theoretical curve (dashed line) is
shifted from the data. The cause of this shift is
the large experimental number of charged 1 and 2 fragments. By simply
removing the charge 1 and 2 fragments from the cumulative charge
distribution, good agreement is achieved.
With the charge $z = 1$
and $z = 2$ fragments included, agreement can be restored by using
two $x$'s as shown in Fig.7a (dotted line),
0.66 of $x = 0.5$ and 0.34 of $x = 50$, Eq.(\ref{ptwox}).
The cumulative mass shows a uniform staircase behavior in the middle
region (see the dotted line in Fig.7b).
Fig.7b shows an interesting property of the data.
Specifically, by interchanging the horizontal and vertical
axes, the appearance of an error function like property of the
cumulative mass distribution can be seen.
An error function is a cumulative function of a Gaussian function;
\begin{eqnarray}
 s = a~ {\rm erf}(x) = a \frac{2}{\sqrt{\pi}} \int_0^x e^{- x^2} ~dx
    \hspace{1cm} {\rm with} \hspace{1cm} x = b (M(s) - M_0) .  \label{erorfnt}
\end{eqnarray}
An error function fit to the data with $a = 79$, $b = 1.6/(79 - M_0)$,
and $M_0 = 25$ is shown in Fig.7b by a dash-dotted line.
The error function fit to the $x$ model for small $x$  is also shown in
Fig.7a by a dash-dotted line ($a = 79$, $b = 2.4/79$, and $M_0 = 0$).
These points related to the error function will be taken up
in another paper.

We next investigate the power moments and generalized Renyi entropies
associated with the data of Waddington and Freier.
Note here that the factorial moments of the multiplicity distribution
$n_z$ for this Au data has been analyzed by other groups
\cite{ploszaj,fracnuc2,xmix} in terms of the intermittency.
As we can deduce from Figs.6 and 7 (cf., discussions at the end of
the paragraph preceeding Eq.(\ref{fqltld})), the factorial moments of $n_z$
essentially comes from the first bin except the case of bin size 1
\cite{ploszaj,fracnuc2}.
On the other hand, since the probability $P_Z(z) = z n_z/Z$ is less than 1,
we can not consider a factorial moments of the probability $P_Z(z)$.
Thus we consider here only the power moments of the probability
distributions; $\sum_J <(zn_z/Z)^q>$ and $\sum_J <zn_z/Z>^q$.
Fig.8a shows the results which exhibit a power law behavior,
i.e., the size $L$ independence of $D_q(L)$; $D_q(L) = 0.17$, $0.22/2$,
$0.24/3$, and $0.25/4$ for the data for $q = 2$, 3, 4, and 5 respectively.
The $L$ independence of $D_q$ indicates the scale invariance of the data.
The $q$ dependence of $D_q$ originates from the multiprocess property of
the data (mixture of $x$'s, low temperature and high $T$);
$(q-1)D_q$ are approximately independent of $q$.
In contrast to this, for the power moments of the mean distribution (Fig.8b)
the generalized dimension is independent of $q$ ($D_q = 0.89$, 0.90,
0.88, and 0.87 for $q = 2$, 3, 4, and 5 respectively) and
shows an $L$ dependence originating from the even-odd oscillating
behavior of $<n_k>$ (bending of the power moment at $L = 2$ in Fig.8b).
Fig.9 shows the fit of the power moments for $q = 2$ and 3 with
our $x$ model. This figure shows that the data is fit better with a
mixture of at least three different values of $x$ for the power moment.
However, for the mean distribution (Figs.6 and 7), only two $x$ values
were enough for a good fit.

Another interesting feature of the data relates to the probability
distributions associated with a fixed $z$. Specifically, what
fraction of the events have $n_z = 0$, $n_z = 1$, ... for a given
charge $z$.
Fig.10 gives a plot of the probability distributions associated with
the charge $z = 1$ (solid circle) and charge $z = 2$ case (diamond).
These distributions are not
Poisson distributions. They are best described by an exponential
behavior given by
\begin{equation}
p(n) ~=~ e^{-n/a}/a  \label{expft}
\end{equation}
\noindent where $a = 13$ and 5 for $z = 1$ and 2 respectively.
This non Poissonian behavior and the mix of two $x$ values
are responsible for the difference between the data and the $x$ model.

Finally, the exponential distribution $p(n)$ given above can be
generated from a Poisson distribution by using a technique used by
R.A. Fisher \cite{azm1} in his theory of species diversity. The interrelation
between models of nuclear fragmentation and theories of species
diversity and evolutionary population genetics in biology was
discussed in Ref.\cite{azm2}. Specifically, in R.A. Fisher's discussion, a
Poisson distribution $p(m,n)~=~ m^ne^{-m}/n!,~ {\rm with}~ m ~=~
<n>$, was randomized with an Eulerian or $\chi^2$ density
distribution $u(m,r,k)dm ~=~ r^{-k}m^{k-1}e^{-m/r} dm/(k-1)!$. The
resulting density
\begin{equation}
p(n,r,k) ~=~ \frac{(k+n-1)!}{(k-1)!n!} ~ \left( \frac{r}{1+r} \right)^n ~
\left(\frac{1}{1+r}\right)^k
\end{equation}
is a negative binomial distribution with parameters $r$ and $k$. For the
special choice $k=1$,
\begin{equation}
p(n,r,k=1) ~=~ \left( \frac{r}{1+r} \right)^n \left(\frac{1}{1+r}\right)
\end{equation}
which can be rewritten in an exponential form with $r/(1+r) = e^{\xi}$.
A distribution in $m~=~<n>$ corresponds to a distribution
in the parameter $x$, since the mean cluster distributions are
related to $x$ (c.f., Subsection 2.1).
The importance of the negative binomial in
understanding intermittency has been stressed by others
\cite{azm3,azm4,azm5} for a
different experimental observation related to rapidity distributions.
Here, we points out the appearance of the negative binomial in a
context connected with the distribution of clusters for small cluster
sizes in this preliminary study of the experimental data.

\section{Conclusion and Summary}

\hspace*{30pt}
In this paper we stressed the importance and properties of the mass
distribution associated with a collision event besides the cluster
distribution used by others. Specifically, we used the fraction of the
mass in clusters of a given size to obtain power moments of
a probability distribution
which was then used to calculate generalized Renyi entropies. This
fraction sums to one event-by-event when summed over all cluster sizes
in each event. In turn, the generalized Renyi entropies are studied
for their scaling properties. The generalized Renyi entropies can also
be used to obtain generalized dimensions from such scaling properties.
These dimensions include not only the Hausdorff covering dimension,
but also the information dimension associated with the Shannon
information content of the probability distribution and the
correlation dimension related to the correlation function of a
probability distribution.

A simple exactly solvable statistical mechanics model of nuclear
fragmentation was used to investigate issues related to
self-similarity, intermittency, and scaling behavior in nuclear
collisions. Correlations and fluctuations were related to the scale
invariance and self similarity of the distribution of fragments. Our
theoretical studies show that the power moments have a power law
scale-invariant behavior in one regime of the thermodynamic variables
of the system, and an exponential behavior with changes in length
scales in another regime of these variables. The thermodynamic
variables are contained in the variable $x$ discussed in the text.

In our studies we also considered the factorial moments of the
distribution of cluster sizes which have also been discussed by several
other groups as noted in Section 2.2. These factorial moments have
been shown by Bialas and Peschanski to be a very useful way of
studying non-Poissonian fluctuations. The theoretical model developed
in this paper in its simplest form (single value of the thermodynamic
parameter $x$) is shown to have a probability distribution of a
given cluster size which is very near Poissonian for reasonably large
values of the initial number of nucleons. To produce non-Poissonian
fluctuations a mixture of different values of $x$ has to be taken.

The staircase properties of a cumulative mass distribution was also
given. The cumulative mass distribution is somewhat analogous to the
cumulative level density distribution used in the discussion of random
matrix models of nuclear level distributions. An error function like
property of the cumulative mass distribution was noted in the results
developed in this paper.

A preliminary discussion of a set of experimental data is presented.
Predictions based on our theoretical model are compared with some
features of this data. Good agreement is obtained for some aspects of
the data such as the overall behavior of inclusive distributions and
the cumulative mass distribution. However, the simple model with a
single value of $x$ does not account for the non-Poissonian
probability distribution seen in the data. We show that this
probability distribution has an exponential behavior for charged
fragments with small charges, i.e., $z=1$, 2, 3, rather than a
Poisson distribution.
Power moments show that the data has a scale invariance, i.e.,
a bin size $L$ independence of the generalized dimension $D_q$.
On the other hand, the multistep fragmentation process of the
emulsion data is reflected by the $q$ dependence of $D_q$ and
a mixture of $x$'s.

This work was supported by National Science Foundation grant number
89-03457.
We thank Prof. Waddington for giving us permission to use his
unpublished data.

\vspace{1cm}

\setlength{\baselineskip}{5ex}
\appendix{{\large\bf APPENDIX}}
\renewcommand{\theequation}{\thesection.\arabic{equation}}
\setcounter{equation}{0}

\section{Factorial Expansion}
\setcounter{equation}{0}

\hspace*{30pt}
Similar to the multinomial expansion
\begin{eqnarray}
 && (x_1 + x_2 + \cdots + x_i + \cdots + x_N)^M
     = \sum_{\{n_i\}_M} M! \prod_{i=1}^N \left[\frac{x_i^{n_i}}{n_i!}\right]
         \nonumber \\
 && M = \sum_{i=1}^N n_i ,  \label{multexp}
\end{eqnarray}
we have a factorial expansion
\begin{eqnarray}
 \frac{\Gamma(x_1 + x_2 + \cdots + x_i + \cdots + x_N + 1)}
      {\Gamma(x_1 + x_2 + \cdots + x_i + \cdots + x_N + 1 - q)}
  = \sum_{\{q_i\}_q} q! \prod_{i=1}^N \left[\frac{1}{q_i!}
        \frac{\Gamma(x_i+1)}{\Gamma(x_i+1-q_i)} \right] ,  \label{factexp}
\end{eqnarray}
where
\begin{eqnarray}
 \sum_{i=1}^N q_i = q .
\end{eqnarray}
It is easy to show this expansion Eq.(\ref{factexp}) of a factorial.
For the $q = 0$ case, we have only one possible partition $\{q_i\}_q$
of $q = 0$ with $q_i = 0$ for all $i$ and thus both the
left and right hand sides are one in Eq.(\ref{factexp}).
For the $q = 1$ case, only one of the $q_i$'s is 1 and the others are zero
for a partition $\{q_i\}$ and thus it is easy to show that Eq.(\ref{factexp})
is satisfied, i.e., both sides are $x_1 + x_2 + \cdots + x_N$.

If we assume Eq.(\ref{factexp}) is true for $q-1$, then for $q$,
we have
\begin{eqnarray}
  & & \hspace{-1.7cm}
 \frac{\Gamma(x_1 + \cdots + x_N + 1)}{\Gamma(x_1 + \cdots + x_N + 1 - q)}
           \nonumber \\
  &=& \frac{\Gamma(x_1 + \cdots + x_N+1)}{\Gamma(x_1 + \cdots + x_N+1-(q-1))}
        (x_1 + \cdots + x_N + 1 - q)         \nonumber \\
  &=& \sum_{\{q_i\}_{(q-1)}} (q-1)! \prod_{j=1}^N \left[ \frac{1}{q_j!}
        \frac{\Gamma(x_j + 1)}{\Gamma(x_j + 1 - q_j)} \right]
        \left[ \sum_{k=1}^N \left(x_k - q_k\right) \right]_{\sum_i q_i = q-1}
                   \nonumber \\
  &=& \sum_{\{q_i\}_{(q-1)}} \sum_{k=1}^N \frac{(q_k + 1)}{q} q!
   \left[\frac{1}{(q_k+1)!} \frac{\Gamma(x_k+1)}{\Gamma(x_k+1-(q_k+1))}\right]
      \prod_{j \ne k} \left[\frac{1}{q_j!}
        \frac{\Gamma(x_j+1)}{\Gamma(x_j+1-q_j)} \right]  \nonumber \\
  &=& \sum_{\{q_i\}_q} \sum_{k=1}^N \frac{q_k}{q} q!
    \prod_{j=1}^N \left[\frac{1}{q_j!}
         \frac{\Gamma(x_j+1)}{\Gamma(x_j+1-q_j)}\right]  \nonumber \\
  &=& \sum_{\{q_i\}_q} q! \prod_{j=1}^N \left[ \frac{1}{q_j!}
        \frac{\Gamma(x_j+1)}{\Gamma(x_j + 1 - q_j)} \right] .
\end{eqnarray}
Thus if Eq.(\ref{factexp}) is satisfied for $q-1$, then it is satisfied
for $q$ case also. Then by induction Eq.(\ref{factexp}) is
satisfied for any $q$.

Similarly, we can also show that the expansion of the upward factorial
is given by
\begin{eqnarray}
 \frac{\Gamma(x_1 + x_2 + \cdots + x_i + \cdots + x_N + q)}
      {\Gamma(x_1 + x_2 + \cdots + x_i + \cdots + x_N)}
  = \sum_{\{q_i\}_q} q! \prod_{i=1}^N \left[\frac{1}{q_i!}
        \frac{\Gamma(x_i+q_i)}{\Gamma(x_i)} \right] ,  \label{factxup}
\end{eqnarray}
with $\sum_{i=1}^N q_i = q$ .
Eqs.(\ref{factexp}) and (\ref{factxup}) are the same for $q = 0$ and 1.
Notice the different form of the $\Gamma$ function between
Eq.(\ref{factexp}) and Eq.(\ref{factxup}).

\section{Recurrence Relation}
\setcounter{equation}{0}

\hspace*{30pt}
Using the grand canonical partition function, we can
obtain a recurrence function for $Q_A(x)$ and
$D_n^{(1,2,...,r)}(x)$ in the $x$ model.
In the $x$-model the generating function for $Q_A(x)$ is
the grand canonical partition function
\begin{eqnarray}
 Q(u,x) = \exp\left[x\left(u + \frac{u^2}{2} + \frac{u^3}{3}
             + \cdots\right)\right]
        = \frac{1}{(1 - u)^x} = \sum_{A=0}^\infty Q_A(x) \frac{u^A}{A!} .
\end{eqnarray}
Differentiating both side of this equation with respect to $u$,
we get the recurrence relation for $Q_A(x)$ which is
$Q_{A+1}(x) = (x + A) Q_A(x)$.
Defining
\begin{eqnarray}
 \tilde Q_A(\vec x) &=& \frac{Q_A(\vec x)}{A!}   \nonumber \\
 \tilde D_n^{(r)}(\vec x) &=& \frac{D_n^{(r)}(\vec x)}{n!} ,   \label{tildqd}
\end{eqnarray}
the recurrence relation for $Q_A(x)$ becomes
$(A+1) \tilde Q_{A+1}(x) = (x+A) \tilde Q_A(x)$.

For the case of $x_r = x y$ for $r$-mers and $x_i = x$ otherwise,
the grand canonical partition function
becomes
\begin{eqnarray}
 Q^{(r)}(u,x,y) &=& \exp\left[x\left(u + \frac{u^2}{2} + \frac{u^3}{3}
             + \cdots\right) - (1-y)x \frac{u^r}{r}\right] \nonumber \\
       &=& \frac{1}{(1 - u)^x} \exp\left[-(1-y)x \frac{u^r}{r}\right]
    = \sum_{n=0}^\infty D_n^{(r)}(x,y) \frac{u^n}{n!} .
\end{eqnarray}
This gives
\begin{eqnarray}
 \frac{d}{d u} Q^{(r)}(u,x,y) &=& \frac{x}{(1 - u)^{x+1}}
              \exp\left[-(1-y)x \frac{u^r}{r}\right]   \nonumber \\
   & & \hspace{1.5cm}  - \frac{1}{(1 - u)^x} (1-y)x u^{r-1}
                \exp\left[-(1-y)x\frac{u^r}{r}\right]      \nonumber \\
   &=& \left[ \frac{x}{1 - u} - (1-y)x u^{r-1} \right] Q^{(r)}(u,x,y)
                   \nonumber \\
   &=& \sum_{n=1}^\infty D_n^{(r)}(x,y) \frac{u^{n-1}}{(n-1)!} .
\end{eqnarray}
Thus we have
\begin{eqnarray}
 & &(1 - u) \sum_{n=1}^\infty D_n^{(r)}(x,y) \frac{u^{(n-1)}}{(n-1)!}
            \nonumber  \\   & & \hspace{3cm}
    = x \left[ 1 - (1-y) u^{r-1} + (1-y) u^r \right]
           \sum_{n=0}^\infty D_n^{(r)}(x,y) \frac{u^n}{n!} .
\end{eqnarray}
The terms of $n$'th order in $u$ give
\begin{eqnarray}
 D_{n+1}^{(r)}(x,y) - n D_n^{(r)}(x,y) &=& x D_n^{(r)}(x,y)
         - (1-y) x \tilde n^{r-1} D_{n-r+1}^{(r)}(x,y)  \nonumber \\
     & & \hspace{1cm}  + (1-y) x \tilde n^r D_{n-r}^{(r)}(x,y) ,
\end{eqnarray}
where $\tilde n^q$ is defined by Eq.(\ref{factniq}).
Finally the recurrence relation is given by
\begin{eqnarray}
 (n+1) \tilde D_{n+1}^{(r)}(x,y) &=& (n + x) \tilde D_n^{(r)}(x,y)
                  - (1-y) x \tilde D_{n-r+1}^{(r)}(x,y)  \nonumber \\
  & &  \hspace{1cm}  + (1-y) x \tilde D_{n-r}^{(r)}(x,y)  \label{drrec}
\end{eqnarray}
where $\tilde D_n^{(r)}(x,y)$ defined by Eq.(\ref{tildqd}).
When $y = 0$, this Eq.(\ref{drrec}) reduces to Eq.(\ref{drrecc}),
the reccurence relation for $r$-mer in $x$ model.

When $x_i = xy$ for $i = 1$, 2, ..., $r$-mers, and $x_i = x$ otherwise,
the grand canonical partition function is
\begin{eqnarray}
 Q^{(1,2,...,r)}(u,x,y)
   &=& \frac{1}{(1-u)^x} \exp\left[-(1-y)x(u + \frac{u^2}{2}
            + \cdots + \frac{u^r}{r})\right]   \nonumber \\
   &=& \sum_{n=0}^\infty D_n^{(1,2,...,r)}(x,y) \frac{u^n}{n!} . \label{q1tor}
\end{eqnarray}
Eq.(\ref{q1tor}) can be used to obtain
\begin{eqnarray}
 (1 - u) \frac{d}{d u} Q^{(1,2,...,r)}(u,x,y)
  &=& \left[ x - (1-y)x (1-u^r) \right] \sum_{n=0}^\infty
          \tilde D_n^{(1,2,...r)}(x,y) u^n  \nonumber \\
  &=& (1 - u) \sum_{n=1}^\infty \tilde D_n^{(1,2,...,r)}(x,y) u^{n-1} .
\end{eqnarray}
Thus the recurrence relation becomes
\begin{eqnarray}
(n+1)\tilde D_{n+1}^{(1,2,...,r)}(x,y) = (n + yx) \tilde D_n^{(1,2,...,r)}(x,y)
        + (1-y) x \tilde D_{n-r}^{(1,2,...,r)}(x,y) .   \label{d12rrec}
\end{eqnarray}

When $x_i = xy$ for $r = s$, $s + 1$, ..., $r$-mers, and $x_i = x$
otherwise, the grand canonical partition function is
\begin{eqnarray}
 Q^{(s,s+1,...,r)}(u,x,y)
   &=& \frac{1}{(1-u)^x} \exp\left[-(1-y)x(\frac{u^s}{s} + \frac{u^{s+1}}{s+1}
             + \cdots + \frac{u^r}{r})\right]  \nonumber \\
   &=& \sum_{n=0}^\infty \tilde D_n^{(s,s+1,...,r)}(x,y) u^n .
\end{eqnarray}
This gives
\begin{eqnarray}
 (1 - u) \frac{d}{d u} Q^{(s,s+1,...,r)}(u,x,y)
  &=& \left[ x - (1-y)x u^{s-1}(1 - u^{r-s+1}) \right]
        \sum_{n=0}^\infty \tilde D_n^{(s,s+1,...r)}(x,y) u^n  \nonumber \\
  &=& (1 - u) \sum_{n=1}^\infty \tilde D_n^{(s,s+1,...,r)}(x,y) u^{n-1} .
\end{eqnarray}
Thus the recurrence relation becomes
\begin{eqnarray}
 (n+1) \tilde D_{n+1}^{(s,s+1,...,r)}(x,y) &=& (n + x) D_n^{(s,s+1,...,r)}(x,y)
        - (1-y) x \tilde D_{n-s+1}^{(s,s+1,...,r)}(x,y)
                      \nonumber \\
   & & \hspace{2cm} + (1-y) x \tilde D_{n-r}^{(s,s+1,...,r)}(x,y).
                      \label{dsrrec}
\end{eqnarray}
Also $D_n^{(s,s+1,...,r)}(\vec x) = Q_n(\vec x)$
for $n < s$.

\section{Factorial Moments}
\setcounter{equation}{0}

\hspace*{30pt}
The factorial moments for each size is
\begin{eqnarray}
 \left<\frac{n_i!}{(n_i-q)!}\right>
   = \sum_{\{n_i\}} \frac{n_i!}{(n_i-q)!} P_A(\vec n)
   = \sum_{n_i=0}^\infty \frac{n_i!}{(n_i-q)!} P(n_i)
   = \left<\frac{n_i!}{(n_i-q)!}\right>_i
\end{eqnarray}
with
\begin{eqnarray}
 P(n_i) &=& \sum_{\{n_j\}_{n_i}} P_A(\vec n) , \nonumber \\
 P(\vec n_i(L)) &=& \sum_{\{n_j\}_{n_i,n_{i+1},...,n_{i+L-1}}} P_A(\vec n)
\end{eqnarray}
where $\vec n_i(L) = (n_i, n_{i+1}, n_{i+2}, ..., n_{i+L-1})$.
If the bin size is $L$, then, using Eq.(\ref{factexp}) for $N_i(L) =
\sum_{j=1}^L n_{i+j-1}$,
\begin{eqnarray}
 \left<\frac{N_i(L)!}{(N_i(L)-q)!}\right>
   &=& \sum_{\{n_i\}} \frac{\Gamma(n_i + n_{i+1} + \cdots + n_{i+L-1} +1)}
                       {\Gamma(n_i + n_{i+1} + \cdots + n_{i+L-1} +1 - q)}
              P_A(\vec n)   \nonumber \\
   &=& \sum_{\{n_i\}} \sum_{\{q_j\}_q} q! \prod_{j=1}^L \left[\frac{1}{q_j!}
        \frac{\Gamma(n_{i+j-1} + 1)}{\Gamma(n_{i+j-1} + 1 - q_j)} \right]
          P_A(\vec n)     \nonumber \\
   &=& \sum_{\{q_j\}_q} q! \sum_{\{n_i\}} \prod_{j=1}^L \left[\frac{1}{q_j!}
        \frac{\Gamma(n_{i+j-1} + 1)}{\Gamma(n_{i+j-1} + 1 - q_j)}
          \right] P_A(\vec n)       \nonumber \\
   &=& \sum_{\{q_j\}_q} q! \sum_{\{\vec n_i(L)\}}
        \prod_{j=1}^L \left[\frac{1}{q_j!}
        \frac{\Gamma(n_{i+j-1} + 1)}{\Gamma(n_{i+j-1} + 1 - q_j)}
          \right] P(\vec n_i(L))       \nonumber \\
   &=& \sum_{\{q_j\}_q} q! \left[ \prod_{j=1}^L \frac{1}{q_j!}\right]
        \left< \prod_{j=1}^L \left[\frac{\Gamma(n_{i+j-1} + 1)}
          {\Gamma(n_{i+j-1} + 1 - q_j)} \right] \right> .
\end{eqnarray}
Using the notation of Eq.(\ref{factniq}),
we have
\begin{eqnarray}
 \left<\tilde N_i^q(L)\right>
   &=& \sum_{\{q_j\}_q} q! \left[ \prod_{j=1}^L \frac{1}{q_j!}\right]
        \left< \prod_{j=1}^L \tilde n_{i+j-1}^{q_j} \right>   \nonumber  \\
   &=& \sum_{i_1=i}^{i+L-1} \sum_{i_2=i}^{i+L-1} \cdots \sum_{i_q=i}^{i+L-1}
        \left< \prod_{j=1}^q \tilde n_{i_j}^{q_j} \right> .
      \label{factbin}
\end{eqnarray}
On the other hand, due to the multinomial expansion Eq.(\ref{multexp}),
we also have
\begin{eqnarray}
 \left<N_i^q(L)\right>
   &=& \sum_{\{q_j\}_q} q! \left[ \prod_{j=1}^L \frac{1}{q_j!}\right]
        \left< \prod_{j=1}^L n_{i+j-1}^{q_j} \right>   \nonumber  \\
   &=& \sum_{i_1=i}^{i+L-1} \sum_{i_2=i}^{i+L-1} \cdots \sum_{i_q=i}^{i+L-1}
        \left< \prod_{j=1}^q n_{i_j}^{q_j} \right> .
      \label{powrbin}
\end{eqnarray}

For $Q_A(\vec x)$, from Eqs.(\ref{factbin}) and (\ref{nkcorr}),
\begin{eqnarray}
 {\cal N}_J^q(\vec x,L) &\equiv& <\tilde N_J^q(L)> \nonumber \\
      &=& \sum_{\{q_j\}_q} q!
                \prod_{j=1}^L \left(\frac{1}{q_j!} \left[
                 \frac{x_{(J-1)L + j}}{(J-1)L + j}\right]^{q_j}\right)
                \frac{A!}{Q_A(\vec x)}
          \nonumber  \\  & &  \hspace{2cm} \times ~
               \frac{Q_{A - \left\{\sum_{j=1}^L ((J-1)L + j) q_j
                       \right\}}(\vec x)}
             {(A - \left\{\sum_{j=1}^L ((J-1)L + j) q_j\right\})!}
                     \nonumber \\
          &=& \sum_{j_1 = (J-1)L+1}^{JL} \sum_{j_2 = (J-1)L+1}^{JL}
                  \cdots \sum_{j_q = (J-1)L+1}^{JL}
              \left[\frac{x_{j_1}}{j_1} \frac{x_{j_2}}{j_2} \cdots
                    \frac{x_{j_q}}{j_q} \right]
          \nonumber  \\  & &  \hspace{2cm} \times ~
              \frac{A!}{Q_A(\vec x)}
           \frac{Q_{A - \{j_1 + j_2 + \cdots + j_q\}}(\vec x)}
                  {(A - \{j_1 + j_2 + \cdots + j_q\})!}
                     \label{fqlnumap}
\end{eqnarray}
and, from Eqs.(\ref{multexp}) and (\ref{nkcorr}),
\begin{eqnarray}
 {\cal D}_J^q(\vec x,L) &\equiv& <N_J(L)>^q  \nonumber \\
         &=& \sum_{\{q_j\}_q} q!
                \prod_{j=1}^L \left( \frac{1}{q_j!}
                  \left[\frac{x_{(J-1)L + j}}{(J-1)L + j}
                  \frac{A!}{Q_A(\vec x)}
                  \frac{Q_{A - \{(J-1)L + j\}} (\vec x)}{(A-\{(J-1)L + j\})!}
                 \right]^{q_j} \right)     \nonumber \\
          &=& \sum_{j_1 = (J-1)L+1}^{JL} \sum_{j_2 = (J-1)L+1}^{JL}
                  \cdots \sum_{j_q = (J-1)L+1}^{JL}
              \left[\frac{x_{j_1}}{j_1} \frac{x_{j_2}}{j_2} \cdots
                    \frac{x_{j_q}}{j_q} \right]
     \left[\frac{A!}{Q_A(\vec x)} \frac{Q_{A - j_1}(\vec x)}{(A-j_1)!}\right]
          \nonumber  \\  & &  \hspace{2cm} \times ~
     \left[\frac{A!}{Q_A(\vec x)} \frac{Q_{A - j_2}(\vec x)}{(A-j_2)!}\right]
                   \cdots
     \left[\frac{A!}{Q_A(\vec x)} \frac{Q_{A - j_q}(\vec x)}{(A-j_q)!}\right]
                   \nonumber \\
          &=& \left[ \sum_{j = (J-1)L+1}^{JL}
                  \left(\frac{x_j}{j}\right) \frac{A!}{Q_A(\vec x)}
                  \frac{Q_{A - j}(\vec x)}{(A - j)!} \right]^q
                     \label{fqldenap}
\end{eqnarray}

\vspace{1cm}
%\pagebreak

\pagebreak

{\bf Table 1:} Bin size $L$ dependence of a scaled factorial moments
$F_q(L)$ of Eq.(\ref{fqldef}) for various values of $x$; $F_q(L)
\propto L^{\alpha (q)} = L^{(q-1) \alpha} = L^{q \alpha'}$ or
$F_q(L) \propto e^{-L/\xi(q)} = e^{-L (q-1)^2/\xi}$.
Also shown is the $L$ dependence of $\tilde F_q(L)$ given by
Eq.(\ref{fqltld});
$\tilde F_q(L) \propto L^{\beta(q)} = L^{(q-1) \beta} = L^{q \beta'}$.

\begin{center}
\vspace{0.5cm}
\smallskip

\begin{tabular}{cc|ccc|cc|ccc}
\hline &   &      &       &      &  &  &      &      &      \\
 $x$ & $q$  &  $\alpha(q)$  &  $\alpha$  &  $\alpha'$  &  $\xi(q)$  &  $\xi$
           &   $\beta(q)$   &  $\beta$   &  $\beta'$  \\
\hline &   &      &       &      &  &  &      &      &      \\
 0.001 & 2 & 1.96 & 1.96  & 0.98 &  &  & 1.96 & 1.96 & 0.98 \\
       & 3 & 2.49 & 1.24  & 0.83 &  &  & 2.49 & 1.25 & 0.83 \\
       & 4 & 2.91 & 0.97  & 0.73 &  &  & 2.92 & 0.97 & 0.73 \\
 0.01  & 2 & 1.40 & 1.40  & 0.70 &  &  & 1.44 & 1.44 & 0.72 \\
       & 3 & 1.90 & 0.95  & 0.63 &  &  & 1.97 & 0.98 & 0.66 \\
       & 4 & 2.30 & 0.77  & 0.58 &  &  & 2.39 & 0.80 & 0.60 \\
 0.1   & 2 & 0.71 & 0.71  & 0.35 &  &  & 0.95 & 0.95 & 0.48 \\
       & 3 & 1.19 & 0.60  & 0.40 &  &  & 1.46 & 0.73 & 0.49 \\
       & 4 & 1.63 & 0.54  & 0.41 &  &  & 1.87 & 0.62 & 0.47 \\
 0.3   & 2 &   &   &   &      &        & 0.86 & 0.86 & 0.43 \\
       & 3 &   &   &   &      &        & 1.55 & 0.78 & 0.52 \\
       & 4 &   &   &   &      &        & 2.17 & 0.72 & 0.54 \\
 0.5   & 2 &   &   &   &      &        & 0.86 & 0.86 & 0.52 \\
       & 3 &   &   &   &      &        & 1.55 & 0.77 & 0.52 \\
       & 4 &   &   &   &      &        & 2.16 & 0.72 & 0.54 \\
 1.0   & 2 &   &   &   &      &        & 0.85 & 0.85 & 0.43 \\
       & 3 &   &   &   &      &        & 1.54 & 0.77 & 0.51 \\
       & 4 &   &   &   &      &        & 2.15 & 0.72 & 0.54 \\
 10.0  & 2 &   &   &   & 1391 &  1391  & 0.82 & 0.82 & 0.41 \\
       & 3 &   &   &   & 351  &  1405  & 1.49 & 0.74 & 0.50 \\
       & 4 &   &   &   & 163  &  1469  & 2.08 & 0.69 & 0.52 \\
 20.0  & 2 &   &   &   & 1224 &  1224  & 0.71 & 0.71 & 0.35 \\
       & 3 &   &   &   & 312  &  1250  & 1.24 & 0.62 & 0.41 \\
       & 4 &   &   &   & 148  &  1328  & 1.71 & 0.57 & 0.43 \\
 50.0  & 2 &   &   &   & 890  &  890   & \\
       & 3 &   &   &   & 228  &  912   & \\
       & 4 &   &   &   & 108  &  972   & \\
 79.0  & 2 &   &   &   & 992  &  992   & \\
       & 3 &   &   &   & 264  &  1057  & \\
       & 4 &   &   &   & 127  &  1140  & \\
\hline \\
\end{tabular}
\end{center}
\pagebreak

{\bf Table 2:} Power moments ${\cal P}_q(L)$ of Eq.(\ref{pqldef}) of the
mass distribution for various values of $x$ in the $x$ model; ${\cal
P}_q(L) \propto L^{(q-1) D_q}$ or ${\cal P}_q(L) \propto e^{L q / \zeta}$.
$D_q$ is the generalized dimension of Eq.(\ref{sqmeanh}).

\begin{center}
\vspace{0.7cm}

\smallskip

\begin{tabular}{cc|cc}
\hline  &     &               &           \\
  $x$   & $q$ &    $D_q$      &  $\zeta$  \\
\hline  &     &               &           \\
  0.1   &  2  &  0.00039  &  1685  \\
        &  3  &  0.00024  &  1671  \\
  1.0   &  2  &  0.018    &  195   \\
        &  3  &  0.011    &  181   \\
  5.0   &  2  &  0.198    &  14.6  \\
        &  3  &  0.129    &  16.7  \\
  10.0  &  2  &  0.449    &   \\
        &  3  &  0.395    &   \\
  20.0  &  2  &  0.745    &  4.06  \\
        &  3  &  0.751    &  2.98  \\
  50.0  &  2  &  0.836    &   \\
        &  3  &  0.824    &   \\
  79.0  &  2  &  0.800    &   \\
        &  3  &  0.757    &   \\
\hline
\end{tabular}
\end{center}

\pagebreak

{\bf Table 3:} The generalized dimension $D_q$ of Eq.(\ref{sqmeanp})
for a power moment of the mean mass distribution for various $x$.

\begin{center}

\vspace{0.7cm}
\smallskip

\begin{tabular}{c|ccccc}
\hline \\
  $x$   & $q = 1$ &     2   &    3    &    4    &    5    \\
\hline & & \\
  0.01  &  0.083  &  0.023  &  0.017  &  0.015  &  0.014  \\
  0.1   &  0.480  &  0.218  &  0.167  &  0.148  &  0.139  \\
  0.3   &  0.824  &  0.595  &  0.482  &  0.432  &  0.405  \\
  0.5   &  0.942  &  0.849  &  0.756  &  0.695  &  0.656  \\
  1.0   &  1.000  &  1.000  &  1.000  &  1.000  &  1.000  \\
  2.0   &  0.999  &  1.000  &  1.000  &  1.000  &  1.000  \\
  5.0   &  0.999  &  0.999  &  0.998  &  0.998  &  0.998  \\
  10.0  &  0.997  &  0.995  &  0.993  &  0.991  &  0.989  \\
  20.0  &  0.991  &  0.983  &  0.975  &  0.967  &  0.959  \\
  50.0  &  0.957  &  0.919  &  0.886  &  0.857  &  0.833  \\
  79.0  &  0.917  &  0.848  &  0.792  &  0.751  &  0.720  \\
\hline
\end{tabular}
\end{center}

\pagebreak

{\large\bf Figure Captions}
\begin{description}

\item[Fig.1:] Distributions of cluster size $k = 1$, 2, and 4
for $x = 7.6269$ and $A = 200$ (solid lines).
The mean number of monomers is $<n_1> = 7.3823$ for this case.
A Poisson distribution for each cluster size with the same mean
number are indistinguishable in this figure.
Also shown by a dotted line is the monomer distribution for
$A = 26$ with $x = 9.9130$ and having the same mean $<n_1> = 7.3823$.
The dash-dotted line is the monomer distribution
for $x = 0.9 \times 7.6269$
and the dashed line is for $x = 1.1 \times 7.6269$.

\item[Fig.2(a):] Bin size $L$ dependence of the factorial moments
$F_q(L)$ for various $x$ and $q$ in a $\log(F_q(L))$ {\it vs}
$\log(L)$ plot for $A = 79$.
The solid line is for $q = 2$, the dashed line is for $q = 3$,
and the dash-dotted line is for $q = 4$.

\item[Fig.2(b):] Same as Fig.2(a), but in a $\log(F_q(L))$
{\it vs} $L$.

\item[Fig.2(c):] Same as Fig.2(a), but for $\tilde F_q(L)$ of
Eq.(\ref{fqltld}) instead of for $F_q(L)$.

\item[Fig.3:] Staircase properties of the cumulative mass distribution
for $A = 40$. The black square at the bottom represent the existence
of the cluster with the corresponding size.
(a) is for a partition $\vec n = 2^1 3^2 12^1 20^1$ and
(b) is for a partition $\vec n = 1^3 4^2 6^1 7^2 9^1$.

\item[Fig.4(a):] Power moment of the mass distribution $P_A(k,x)$
for various $x$ for $A = 79$.
The figure is a plot of $\log({\cal P}_q(L))$ {\it vs} $\log(L)$.
The solid line is for $q = 2$ and the dashed line is for $q = 3$.

\item[Fig.4(b):] Same as in Fig.4(a), but for $\log({\cal P}_q(L))$
{\it vs} $L$.

\item[Fig.4(c):] Same as in Fig.4(a), but for
the power moment of the mean mass distribution
$<P_A(k,x)>$ instead of $<P_A(k,x)^q>$
for various $x$ for $A = 79$.
Here the dashed, the dash-dotted, the dotted, and the
dash-dot-dot-dotted lines are for $q = 2$, 3, 4, and 5 respectively.
The solid line is for the limit of $q \to 1$, i.e.,
$e^{-S_1(L)}$ where the entropy is given by Eq.(\ref{sqmeanh}).

\item[Fig.5:] Devil's staircase for the Cantor set at various
stages of generation.
The mass bars present at the corresponding generation are
shown at the bottom.

\item[Fig.6:] Cluster size distribution in Au emulsion data
of Ref.\cite{nucdat} (solid circles connected by solid line).
The dashed line is for $x = 0.3$ in the $x$ model
and the dotted line is a mixture of 0.66 of $x = 0.5$ and
0.34 of $x = 50$; Eq.(\ref{ptwox}).
The dash-dotted line is a mixture of 0.72 of $x = 0.5$, 0.15
of $x = 5$, and 0.13 of $x = 79$ which was used in Fig.9. Other data
are discussed in Ref. [12].

\item[Fig.7:] Staircase features of the Au emulsion data (solid line).
(a) Dashed line is for $x = 0.3$ and the dotted line is a mixture
of $x = 0.5$ and $x = 50$ Eq.(\ref{ptwox}) as in Fig.6.
Dash-dot-dot-dotted line is a mixture of 0.72 of $x = 0.5$, 0.15
of $x = 5$, and 0.13 of $x = 79$.
Dash-dotted line is an error function of Eq.(\ref{erorfnt}) with
$a = 79$, $b = 2.4/79$, and $M_0 = 0$.
(b) Dash-dotted line is an error function fit of Eq.(\ref{erorfnt})
with $a = 79$, $b = 1.6/(79 - M_0)$, and $M_0 = 25$. Dotted line
is a straight line fit of $M(s) = (40/79) s + 23$.

\item[Fig.8:] Power moments (a) $<(z n_z/Z)^q>$ and
(b) $<z n_z/Z>^q$. The solid, dashed, dotted, and dash-dotted lines
are for $q = 2$, 3, 4, and 5 respectively.

\item[Fig.9:] Power moments for (a) $q = 2$ and (b) $q = 3$.
The solid lines are the data, the dash-dotted lines are the fit of
Eq.(\ref{ptwox}) with 0.66 of $x = 0.5$ and 0.34 of $x = 50$,
the dashed lines are a mixture of 0.75 of $x = 0.5$ and 0.25
of $x = 20$, and the dotted lines are a mixture of 0.72 of $x = 0.5$,
0.15 of $x = 5$, and 0.13 of $x = 79$.

\item[Fig.10:] The distribution $P(n_z)$ for the charge number
$z = 1$ (solid circle) and $z = 2$ (diamond).
The solid line and the dotted line are the exponential fits
of Eq.(\ref{expft}) with $a = 13$ and 5 respectively.

\end{description}


\begin{thebibliography}{99}
 \bibitem{hanbury}A.Z. Mekjian, Phys. Rev. {\bf 17}, 1051 (1978).
 \bibitem{ploszaj}M. Ploszajczak and A. Tucholski, Nucl. Phys.
   {\bf A523}, 651 (1991).
 \bibitem{fracnuc1}K. Haglin, C. Gale, and S. Das Gupta, McGill Preprint 91-32.
 \bibitem{fracnuc2} D.H.E, Gross, A.R. DeAngelis, H.R. Jaqaman, Pan
Jicai, and R. Heck, Phys. Rev. Lett. {\bf 68}, 146 (1992);\\
A.R. DeAngelis, D.H.E. Gross and R. Heck, Nucl. Phys. {\bf A537}, 606 (1992).
 \bibitem{bialas}A. Bialas and R. Peschanski, Nucl. Phys. {\bf B273},
   703 (1986), {\bf B308}, 857 (1988).
 \bibitem{mandel}B.B. Mandelbrot, J. Fluid Mech. {\bf 62}, 331 (1974).
 \bibitem{turbul}U. Frisch, P. Sulem and M. Meklin, J. Fluid Mech.
{\bf 87}, 719 (1978).
 \bibitem{mekjian}A.Z. Mekjian, Phys. Rev. {\bf C41}, 2103 (1990); \\
   Phys. Rev. Letts. {\bf 64}, 2125 (1990).
 \bibitem{xymodel}S.J. Lee and A.Z. Mekjian, Phys. Lett. {\bf A149},
   7 (1990).
 \bibitem{general}S.J. Lee and A.Z. Mekjian, Phys. Rev. {\bf C45}, 1284 (1992).
 \bibitem{otherx}A.Z. Mekjian and S.J. Lee, Phys. Rev. {\bf A44}, 6294 (1991).
 \bibitem{nucfit}S.J. Lee and A.Z. Mekjian, Phys. Rev. {\bf C45}, 365 (1992).
 \bibitem{xmix}D.H.E. Gross, A. Ecker and A.R. DeAngelis, Preprint,
Hahn-Meitner Institut, Berlin, Mean Values and Fluctuations in the
Fragmentation of Mescopic Systems Using an Exactly Soluable Model with
Strict Particle Conservation and Quantum Symmmetry.
 \bibitem{fractal}M. Schroeder, {\it Fractals, Chaos, Power Laws; Minutes
   from an Infinite Paradise}, (W.H. Freeman and Company, New York, 1991)
 \bibitem{frgmap}S.J. Lee and A.Z. Mekjian, Rutgers preprint \#9226,
Geometry, Scaling and Universality in the Mass Distributions in Heavy
Ions Collisions, submitted for publication.
 \bibitem{grass}P. Grassberger, J. of Statist. Phys. {\bf 26}, 173 (1981).
 \bibitem{barnsley}M. Barnsley, {\it Fractals Everywhere}, (Academic
   Press, Inc., San Diego, 1988).
 \bibitem{nucdat}C.J. Waddington and P.S. Freier, Phys. Rev. {\bf C31}, 888
   (1985).
 \bibitem{azm1} R.A. Fisher, A.S. Corbet, and C.B. Williams, J. Anim.
Ecol. {\bf 12}, 42 (1943).
 \bibitem{azm2}A.Z. Mekjian, Phys. Rev. {\bf A44}, 8361 (1991).
\bibitem{azm3} W. Ochs and J. Wosiek, Phys. Lett. {\bf B232}, 271 (1989).
\bibitem{azm4}P. Carruthers and C.C. Shih, Int. Jour. of Modern Phys.
{\bf A2}, 1447 (1987).
\bibitem{azm5}A. Giovannini and L. van Hove, Z. Phys. {\bf C30}, 391
(1986).

\end{thebibliography}
\end{document}